# Generalized causality constraint based on duality symmetry reveals untapped potential of sound absorption

Sichao Qu, [1,2] Min Yang, [3,*] Sibo Huang, [4] Shuohan Liu, [1] Erqian Dong, [1] Helios Y. Li, [1]
Ping Sheng, [2,†] I. David Abrahams, [2,‡] and Nicholas X. Fang[1,5,§]

*[1]Department of Mechanical Engineering, The University of Hong Kong, Pokfulam Road, Hong Kong, China*
*[2]Department of Applied Mathematics and Theoretical Physics (DAMTP), University of Cambridge, Wilberforce Road, Cambridge CB3 0WA, UK*
*[3]Acoustic Metamaterials Group, Data Technology Hub, No. 5 Chun Cheong Street, Hong Kong, China*
*[4]Department of Electrical Engineering, City University of Hong Kong, Tat Chee Avenue, Kowloon, Hong Kong, China*
*[5]Materials Innovation Institute for Life Sciences and Energy (MILES), HKU-SIRI, Shenzhen, China*

Correspondence to: [*]min@metacoust.com; [†]sheng@ust.hk; [‡]ida20@cam.ac.uk; [§]nicxfang@hku.hk

## Abstract

Causality constraints are known to bind sound absorption to a limit that can only be achieved by optimizing the system bandwidth for a specific material thickness. This limit is defined on the assumption of a one-port system, generally causing duality symmetry to be overlooked. Here, we define a generalized causality constraint of sound absorption by investigating reflection and transmission of a two-port hybrid monopole-dipole resonator. With our theory, we show that the absorption limit is approached by relying on the well-established critical coupling as well as by matching effective compressibility and density. We experimentally show that the designed resonator absorbance follows the duality symmetry condition embodied in the large bandwidth reported. A comparison with a traditional foam liner and other competitive works further validates our findings, confirming an intrinsic connection between duality symmetry and scattering causality. Our results untap unexplored absorption potential in broadband acoustic metamaterials.

## Introduction

Causality, a fundamental principle underpinning the asymmetry of unidirectional time flow, governs the upper bound of absorption integral in passive, linear, and time-invariant (LTI) materials[1,2]. Former investigations have revealed that achieving perfect absorption at discretized frequencies requires critically coupled non-Hermitian resonances at exceptional points[3,4]. The manipulation of zeros and poles on the complex plane of the scattering coefficients follows the balance between the mode density and quality factors of the integrated resonator units, thus achieving broadband impedance matching and high absorption. Subsequent key developments have neared this absorption bound, demonstrated by the broadband absorption, observed in multiple resonance-based metamaterials across acoustic[5], elastic[6] and electromagnetic domains[7]. Studying the absorption bound of causality constraint will guide the design wave functional devices[8-11] with optimal broadband performance in the laboratory and pave the way for large-scale industrial applications[12].

However, current methodologies neglect a critical question: how do fundamental symmetries dictate absorption bound throughout the entire spectrum? We find clues by considering causality constraints together with duality symmetry[13], a fundamental principle embedded in numerous physics equations. For example, the supersymmetric Yang-Mills theory reveals that the strong coupling limit is remarkably equivalent to its weak coupling counterpart, with an interchange of roles between electric and magnetic charges as delineated by Montonen-Olive duality[14,15]. Maxwell equations are also invariant when swapping the electric and magnetic fields[16]. Also, duality symmetry is related to the conservation of the helicity of electromagnetic waves, attributed to the invariant degeneracy of permittivity and permeability across different geometries. In acoustics, the implications of duality symmetry are unclear for this context, due to the longitudinal wave nature, which intrinsically prohibits the helicity[17]. Although the analogy of acoustic duality symmetry has been proposed and investigated in the seminal work[17] on the subject, the relationship between acoustic duality and scattering causality—specifically, its impact on the broadband absorption bandwidth bound of two-port systems—remains unexplored.

In this study, we have revealed the complement of a previously established causality constraint[2,5,9,18] through a duality transformation. This facilitates the derivation of a generalized bound for the absorption integral of passive two-port systems, including both reflection and transmission coefficients[19,20]. Formerly known causality constraints, such as the Rozanov limit[1] and the Fano-Bode bounds[21], along with some derivative versions[9,22-24], are primarily based on one-port configurations, thereby limiting their applicability. Furthermore, we noticed that numerous studies[25-27] interpreted their findings through one-port causality constraints, despite employing two-port experimental setups. In this context, we address this discrepancy by presenting a generalized bound for two-port absorbers, which is at least double that of one-port absorbers. This provides theoretical insights into the evident difficulty of achieving broadband high absorption without a backing condition, compared to the scenarios involving backing, or the equivalent one-port setup.

Through the construction of a surrogate model with Lorentzian dispersion, we have analytically demonstrated the equivalence between fulfilling the proposed bound concerning the preservation of broadband duality symmetry, which encompasses the overlapped monopole and dipole resonances, and satisfying the established critical coupling condition. This condition pertains to the manipulation of zeros to reside on the real frequency axis by adjusting the system dissipation. In addition, we have experimentally demonstrated that broadband duality-symmetric absorption can be achieved through the overlapped first-order monopole-dipole resonances with the matched compressibility and density. Notably, by employing the ratio between the absorption integral and the bound as a Figure of Merit (FOM), we show that the developed dispersion-customized meta-absorber surpasses not only the traditional foam liner but also some previously reported two-port sound-absorbing meta-structures[25-30] with competitive absorption performance.

## Results

### A generalized causality constraint for a two-port absorber

According to Refs.[2,5,31], the original causality constraint for a one-port acoustic absorber is expressed as follows:

$$\int_0^{\infty} -\ln(|S_m(\omega)|^2)\frac{d\omega}{\omega^2} \leq \frac{2\pi D}{c_0}\frac{K_0}{K_{\text{eff}}(0)}, \tag{1}$$

where $D$ represents the sample thickness, $c_0, K_0$ are the sound speed and bulk modulus of the background medium (e.g., air or water), and $K_{\text{eff}}(0)$ denotes the static bulk modulus of the sample. $\frac{K_0}{K_{\text{eff}}(0)}$ is the effective compressibility (the inverse of the normalized bulk modulus). The variable of integration, $d\omega/\omega^2$, is proportional to $d\lambda$ where $\lambda$ denotes the wavelength. According to energy conservation, the absorption $A(\omega)$ is defined as $1-|S_m(\omega)|^2$ where $S_m(\omega)$ is the reflection coefficient. The derivation of Eq. (1) assumes a Neumann backing condition, which implies either $\partial_x p = 0$ or $v = 0$, corresponding to the scattering process depicted in the left plane of Fig. 1(a). Due to mirror symmetry (or anti-symmetry), this boundary condition allows for mapping the pressure field $p$ (or velocity field $v$) from the left plane to the right plane. Consequently, Eq. (1) is applicable to acoustic problems involving symmetrical pressure excitation and $S_m(\omega)$ becomes the amplitude of monopole scattering, also referred as symmetrical coherent perfect absorption[32]. How to reduce $K_{\mathrm{eff}}(0)$ to relax the stringent constraint? This has attracted great research interests with substantial advances in recent publications[9,33-35].

The 1D acoustic equations for harmonic waves can be expressed in a matrix form as: $\partial_x \binom{p}{v} = i\omega \begin{pmatrix} 0 & \rho \\ K^{-1} & 0 \end{pmatrix} \binom{p}{v}$, where $\rho$ and $K$ represent the local mass density and bulk modulus, respectively. In this context, we introduce a duality transformation that mixes the pressure and velocity fields: $\binom{p_\theta}{v_\theta} = \mathcal{D}(\theta) \binom{p}{v} = \begin{pmatrix} \cos\theta & iZ_0 \sin\theta \\ i\sin\theta / Z_0 & \cos\theta \end{pmatrix} \binom{p}{v}$, where $Z_0 = \sqrt{\rho_0 K_0}$ (see the detailed derivation on acoustic duality transformation with arbitrary $\theta$ in Supplementary Note 1). If we apply a duality transformation with $\theta = \frac{\pi}{2}$, we specifically consider the exchange $p \leftrightarrow v$, $\rho_0 \leftrightarrow K_0$, and $\rho_{\mathrm{eff}} \leftrightarrow K_{\mathrm{eff}}$ between Fig. 1(a) and Fig. 1(b). Consequently, the specific impedance is transformed into specific admittance: $Z = \frac{p}{v} \rightarrow Y = Z_{\frac{\pi}{2}} = \frac{v}{p}$. Moreover, in Eq. (1), $S_m = \frac{Z-Z_0}{Z+Z_0}$, which can be transformed into $S_m \rightarrow -S_d$ because $\frac{Z-Z_0}{Z+Z_0} = -\frac{Y-1/Z_0}{Y+1/Z_0}$. Also, the backing condition ($\partial_x p = 0$) in Fig. 1(a) is also replaced by the Dirichlet one ($p = 0$) in Fig. 1(b). Therefore, the causality constraint for monopole scattering becomes the complimentary version for dipole scattering[22]:

$$\int_0^{\infty} -\ln(|S_d(\omega)|^2) \frac{d\omega}{\omega^2} \leq \frac{2\pi D}{c_0} \frac{\rho_{\mathrm{eff}}(0)}{\rho_0}, \tag{2}$$

where $S_d(\omega)$ is either dipole scattering coefficient under anti-symmetrical excitation, or the reflection coefficient of a soft boundary absorber [see Fig. 1(b)]. In particular, in this case one can afford near unity absorption over a reasonably large frequency range for a thin absorber, owing to the generally large ratio of $\frac{\rho_{\mathrm{eff}}(0)}{\rho_0}$[6,22], i.e., the large imaginary part of $\rho_{\mathrm{eff}}(\omega)$ around the dynamic absorbing frequency leads to large static and real-valued $\rho_{\mathrm{eff}}(0)$, allowed by Kramer-Kronig relations.

If one linearly superposes the pressure fields from Fig. 1(a) and Fig. 1(b) to cancel the right-side incident waves and renormalizes the left-side incident field to unity as shown in Fig. 1(c), the reflection and transmission coefficients can be expressed in terms of dipole and monopole scattering coefficients ($S_m$ and $S_d$ respectively):

$$\begin{cases} T(\omega) = \dfrac{S_m(\omega) - S_d(\omega)}{2} \\ R(\omega) = \dfrac{S_m(\omega) + S_d(\omega)}{2} \end{cases}. \tag{3}$$

According to the energy conservation and Eq. (3), $1-A(\omega)=|R(\omega)|^2+|T(\omega)|^2=\frac{|S_m(\omega)|^2+|S_d(\omega)|^2}{2}\geq|S_m(\omega)||S_d(\omega)|$. The last relation is derived from the inequality of arithmetic and geometric means, or more briefly the AM-GM inequality. By converting this relation into a logarithmic function, we obtain $-\ln\big(1-A(\omega)\big)\leq-\ln(|S_m(\omega)|)-\ln(|S_d(\omega)|)$. Based on this, we can combine Eq. (1) and Eq. (2) to yield a generalized causality constraint

$$\int_0^\infty -\ln\big(1-A(\omega)\big)\frac{d\omega}{\omega^2}\leq\frac{1}{2}\int_0^\infty -[\ln(|S_m(\omega)|)+\ln(|S_d(\omega)|)]\frac{d\omega}{\omega^2} \\ \leq\frac{\pi L}{2c_0}\left(\frac{K_0}{K_{\mathrm{eff}}(0)}+\frac{\rho_{\mathrm{eff}}(0)}{\rho_0}\right)=\Gamma, \tag{4}$$

where $L=2D$ (the defined thickness is effectively doubled due to mirror symmetric boundaries) and $\Gamma$ is the established bound for the two-port setup, reflecting the static material properties. Optimal broadband absorption is defined by the proposed 2-port causality constraint. Importantly, the fundamental principle lies on the fact that at long wavelength limit, all 3D scatterings of two-port problems, become specular reflection and plane-wave transmission, defining the effective compressibility and density. Also, the absorption integral is dominated by long wavelength scatterings as per the weighting factor $(1/\omega^2)$. So, the mathematical form of causality constraint inherently embeds the role of duality for two-port 1D systems.

Certain examples in both 2D and 3D scenarios can be simplified to a 1D case, provided that the dimensions of the scattering matrix are equivalent, specifically 2×2 (see Supplementary Figure 7). For instance, a meta-surface arranged in a periodic layout may absorb perfectly at one incident angle, but only undergo total retro-reflection at another angle, referred as extremely asymmetrical absorption[36]. Although the radiation channels are based on plane waves (equivalent to scalar velocity field), there exist vectorial velocity fields in the near-field region, with non-radiative local modes of the scatterer.

**A surrogate model for verifying generalized causality constraint**

Our goal in this section is to numerically verify the above derivation results through a commonly used Lorentzian dispersion models that conforms to the Kramer-Kronig relations[21]. Specifically, our surrogate model adopts effective compressibility and mass density (normalized by those of background fluid):

$$\begin{cases}\dfrac{K_0}{K_{\mathrm{eff}}(\omega)}=\dfrac{\alpha_m\omega_m^2}{\omega_m^2-\omega^2-i\beta_m\omega}\\[2ex] \dfrac{\rho_{\mathrm{eff}}(\omega)}{\rho_0}=\dfrac{\alpha_d\omega_d^2}{\omega_d^2-\omega^2-i\beta_d\omega}\end{cases}, \tag{5}$$

where $\omega_m$ and $\omega_d$ represent the resonance frequencies, $\alpha_m$ and $\alpha_d$ denote the oscillation strengths, and $\beta_m$ and $\beta_d$ are the dissipation factors for monopole and dipole modes, respectively. Analytical scattering coefficients are derived by substituting Eq. (5) into the following formulas for general subwavelength two-port scattering:

$$\begin{cases} S_m = \dfrac{c_0 + i\omega D \dfrac{K_0}{K_{\text{eff}}(\omega)}}{c_0 - i\omega D \dfrac{K_0}{K_{\text{eff}}(\omega)}}, \\ S_d = -\dfrac{c_0 + i\omega D \dfrac{\rho_{\text{eff}}(\omega)}{\rho_0}}{c_0 - i\omega D \dfrac{\rho_{\text{eff}}(\omega)}{\rho_0}} \end{cases} \tag{6}$$

which satisfies the $\frac{\pi}{2}$ duality transformation. The derivation of Eq. (6) is detailed in Supplementary Note 2. By inserting Eq. (5) into Eq. (6), we equate $S_m(\omega) = 0$ and $S_d(\omega) = 0$ (critical coupling conditions for a two-port absorber), because this is also the sufficient and necessary conditions for $A(\omega) = 1$ [as per Eq. (3)]. In this way, we analytically determine the dissipation factors:

$$\begin{cases} \beta_m = \dfrac{\alpha_m \omega_m^2 D}{2c_0} \\ \beta_d = \dfrac{\alpha_d \omega_d^2 D}{2c_0} \end{cases}. \tag{7}$$

With the closed-form $\beta_m$ and $\beta_d$, the critical coupled zeros of $\ln(|S_m(\omega)|)$ and $\ln(|S_d(\omega)|)$ are locked on the real frequency axis. Note that Eq. (7) is also the equivalent sufficient condition[5,37] of achieving the upper bounds of one-port causality constraints [Eq. (1) and Eq. (2)] for a one-port single-mode resonator. For multi-resonant systems, Eq. (7) is not applicable; consequently, the optimal loss must be determined by considering the mode density distribution[8]. A theoretical model based on continuum resonance distribution[37] has been developed for this purpose. In this study, we prefer to utilize the single-mode version of Eq. (5) to gain insights within a more concise framework.

Next, we can simplify Eq. (4) with $\frac{1}{2}\int_0^\infty -[\ln(|S_m(\omega)|) + \ln(|S_d(\omega)|)]\frac{d\omega}{\omega^2} = \Gamma$. Due to the static frequency-dimension boundary condition $S_m(0) = -S_d(0) = 1$, as per Eq. (6), the remaining condition to achieve $\Gamma$, is that

$$S_m(\omega) = -S_d(\omega), \tag{8}$$

which satisfies that $|S_m(\omega)| = |S_d(\omega)|$ from AM-GM inequality. This condition, which dictates a specific amplitude and phase relationship between the monopole and dipole scattering coefficients, is the acoustic equivalent[38,39] of the Kerker condition, known in electromagnetism and leads to a Huygens-source-like radiating or scattering profile. Electromagnetic superabsorption reported in Ref.[40] indeed serves as an optical counterpart to our study. However, superscattering[41] represents a distinct phenomenon that does not necessitate critical coupling; rather, it maximizes backscattering as opposed to minimizing it. This effect can be viewed as an anti-effect to super-absorption, although both phenomena share a common feature: the alignment of two fundamental resonances, albeit with their phases reversed.

By substituting Eq. (6) into Eq. (8), we deduce that ideal duality symmetry mandates the matched effective compressibility and mass density for arbitrary $\omega$:

$$\frac{K_0}{K_{\mathrm{eff}}(\omega)}=\frac{\rho_{\mathrm{eff}}(\omega)}{\rho_0}. \tag{9}$$

In short, if critical coupling conditions are fixed [as per Eq. (7)], approaching the bound $\Gamma$ requires identical scattering amplitudes of the monopole and dipole resonances [as per Eq. (8)], but with a reversed phase ($\pi$), which is equivalent to broadband duality symmetry condition [as per Eq. (9)]. We prove that to saturate this causality-induced limit ($\Gamma$), the overlapping resonance (Kerker) condition must be maintained across frequency, not just at a single point. This is the concept of broadband duality symmetry.

Utilizing a surrogate model for the monopole-dipole resonator, we numerically validate Eq. (4) through the construction of three representative cases (A, B, C), each conforming to the bound defined by the static value $\Gamma=\frac{\pi L}{2c_0}\left(\frac{K_0}{K_{\mathrm{eff}}(0)}+\frac{\rho_{\mathrm{eff}}(0)}{\rho_0}\right)=(\alpha_m+\alpha_d)\frac{\pi L}{2c_0}$, according to Eq. (5). To be consistent, $\Gamma$ of the three cases are set to be identical ($\alpha_m+\alpha_d=2$), and the critical coupling conditions by Eq. (7) are consistently locked in all subsequent numerical cases [see the critical coupled zeros in Fig. 3(b-c)]. The dispersion profiles of case A, B, C are plotted in Fig. 2 with the following details:

A. A single-mode monopole resonator ($\alpha_m=1$ and $\omega_m=\omega_0$ for which the effective bulk modulus $K_{\mathrm{eff}}(\omega)$), and density $\rho_{\mathrm{eff}}(\omega)$ is constant ($\rho_{\mathrm{eff}}(\omega)=\rho_0$). Although a dipole resonator is also possible, its absorption spectrum will not change due to the principle of duality. As illustrated in Fig. 3(a), the absorption at critical coupling for this monopole resonator peaks at 50%, known as the single-mode limit[31,42,43] for the two-port absorption. The substitution of the absorption spectrum into Eq. (4) reveals that the calculated integral ($\int_0^\infty -\ln\left(1-A(\omega)\right)\frac{d\omega}{\omega^2}$) approaches 58%$\Gamma$.

B. A narrowband duality-symmetric monopole-dipole resonator ($\alpha_m=1.8$ and $\alpha_d=0.2$, with $\omega_m=\omega_d=\omega_0$). Here, duality-symmetric absorption reaches a unity absorption only at the resonant frequency $\omega_0$, where $S_m(\omega_0)=S_d(\omega_0)=0$. Upon reapplying Eq. (4), the absorption integral calculated remains at 71%$\Gamma$.

C. A broadband duality-symmetric monopole-dipole resonator. Here, $\alpha_m$ and $\alpha_d$ are set to 1, with $\omega_m$ and $\omega_d$ also equaling $\omega_0$, and identical critical coupled $\beta_m$ and $\beta_d$. Numerical calculations yield an integral value that precisely matches 99%$\Gamma$, verifying the proposed bound.

The above analysis shows the possibility of exploiting the untapped absorption with a monopole-dipole resonator.

The preceding discussion centers on the ideal single monopole and/or single dipole resonances to simplify the analysis and gain insight into the physics. It is, in fact, feasible to experimentally construct an acoustic single-mode monopole or dipole resonator (as per Case A), respectively. For instance, a bubble-screen resonator exemplifies a monopole[44], whereas a membrane-type metamaterial is classified as a dipole[45]. However, for a practical hybrid monopole-dipole resonator, the higher-order modes inevitably appear[31,46]. To take account this effect, we adopt a modified surrogate model:

$$\begin{cases}\dfrac{K_0}{K_{\mathrm{eff}}(\omega)}=\dfrac{\alpha_m\omega_m^2}{\omega_m^2-\omega^2-i\beta_m\omega}+\delta_m(\omega)\\[2ex] \dfrac{\rho_{\mathrm{eff}}(\omega)}{\rho_0}=\dfrac{\alpha_d\omega_d^2}{\omega_d^2-\omega^2-i\beta_d\omega}+\delta_d(\omega)\end{cases}, \tag{10}$$

where $\delta_m$ and $\delta_d$ are introduced to account for the contribution of the higher-order modes which are not well captured in Eq. (5). Seminal examples following the Lorentzian dispersion model are famous for the dynamic negative properties[47,48]. The zero-pole profiles of the scattering coefficients in the modified surrogate model are available in full complex-plane analysis in Supplementary Figure 1.

The existing monopole-dipole resonators[49-51] have only demonstrated *narrowband* duality symmetry in proximity to their resonance frequencies (as per Case B), an effect also referred to as 'degenerate perfect absorption'[50]. However, to the best of our knowledge, the realization of *broadband* duality-symmetric absorption, even for the lowest-order monopole-dipole resonances, has never been explored (as per Case C). The challenge arises from the difficulty in perfectly aligning the two terms in Eq. (10), which means that $\alpha_m \approx \alpha_d$, $\omega_m \approx \omega_d$, $\beta_m \approx \beta_d$ and $\delta_m \approx \delta_d$. Based on the surrogate model, adherence to these conditions is anticipated to exploit the untapped potential with larger absorption integral; although, the effects induced by $\delta_m$ and $\delta_d$ should be also taken into consideration (see the modelling of the higher-order effects in Supplementary Note 2 for the modification of critical coupling condition and the quantification of the absorption integral).

**The design strategy of broadband duality-symmetric absorption**

In our experimental design, to mimic the dispersion in Eq. (10) as a proof of concept, we employ the ventilative metamaterials[52,53] in duct acoustics. Our idea is to approximately match the lowest-order monopole-dipole resonances over the broad band, while we do not expect that $\delta_m(\omega)$ strictly matches $\delta_d(\omega)$. Since the dominant contribution of absorption integral comes from the low-frequency part, we will show that even this compromise strategy is sufficient to exploit considerable untapped absorption. In particular, the proposed monopole-dipole resonator comprises two coupled Helmholtz resonators (CHR) [Fig. 4(a)], which intuitively supports monopole (dipole) mode with symmetric (anti-symmetric) acoustic modes [see Fig. 4(c)]. To simplify our design and simulation, our sample is axisymmetric with the axis of symmetry being the dotted line [see Fig. 4(b)]. According to acoustic homogenization theory[54,55], the static effective properties of the CHR are purely geometry-dependent:

$$\begin{cases} \dfrac{K_0}{\widetilde{K}_{\text{eff}}(0)} \approx \dfrac{V_{\text{air}}}{S_0 L} \\ \dfrac{\tilde{\rho}_{\text{eff}}(0)}{\rho_0} \approx \dfrac{\ell}{L}\dfrac{1}{\phi} + \left(1 - \dfrac{\ell}{L}\right) \end{cases}, \tag{11}$$

where $V_{\text{air}}$ is the resonator's effective air domain volume, the sample thickness is $L$, $S_0$ is the main duct cross-section area, $\phi S_0$ is the cross-section area of the narrowed channel with the perforated ratio $\phi$, $\ell$ is the length of the narrowed channel, and the superscripts '~' distinguish the approximated estimation values from the actual ones. The first line of Eq. (11) is from the Wood's formula[9,55], where the bulk modulus of solid shells is overlooked because the high contrast of the impedance of solid-air interface. We note that the remaining variable $V_{\text{air}}$ should be evaluated in different ways for pure air domain and porous domain ($V_{\text{air}} = V_{\text{air}}^0 + V_{\text{porous}}$). In pure air domain (e.g., the narrowed channel), it is the same as the actual volume $V_{\text{air}}^0$ occupied by air. In porous domain, due to the comparable scale of micro-structures and viscous boundary layer thickness, the porous material absorption should be treated as isothermal process, according to Ge et al[9] and then, $V_{\text{porous}} = \gamma_{\text{air}} V_{\text{res}}^0 \varphi$ where $V_{\text{res}}^0$ is the resonator's cavity volume without porous materials, the porosity $\varphi$ excludes the porous skeleton volume and $\gamma_{\text{air}}$ is the adiabatic index of air ($\gamma_{\text{air}} = 1.4$). In addition, the second line of Eq. (11) is adapted from the analytical impedance of a micro-perforated plate (MPP)[54], with the end correction effects omitted for simplification. Additionally, thermoviscous dissipation is disregarded due to the boundary layer thickness being much smaller than the channel diameter.

At the static limit ($\omega \to 0$), the broadband duality-symmetric absorption requires that $\frac{K_0}{K_{\text{eff}}(0)} \approx \frac{\rho_{\text{eff}}(0)}{\rho_0}$. First, we set the following parameters; $\phi = 0.25$, $L = 2.5$ cm, and an initial length $\ell$ (randomly selected), to determine $\tilde{\rho}_{\text{eff}}(0)$. Next, we install two identical Helmholtz resonators and monotonously increase their Helmholtz cavity volume until the condition $\frac{K_0}{\tilde{K}_{\text{eff}}(0)} = \frac{\tilde{\rho}_{\text{eff}}(0)}{\rho_0}$ is satisfied as per Eq. (11). This approach works because the left-hand term is effectively adjusted by the resonator's air volume, which is primarily influenced by the cavity volume.

By using the finite-element-method (FEM) simulation, we can further accurately extract the dynamic $K_{\text{eff}}(\omega)$ and $\rho_{\text{eff}}(\omega)$ from the initial structure's reflection $R(\omega)$ and transmission coefficients $T(\omega)$, according to the algorithm in Ref.[56]. The spectral profiles of the extracted $K_{\text{eff}}(\omega)$ and $\rho_{\text{eff}}(\omega)$ indicate their respective ($\alpha_m$, $\omega_m$, $\beta_m$) and ($\alpha_d$, $\omega_d$, $\beta_d$), as we have seen in Fig. 3. See Methods for the details of the FEM simulation. To tune $\omega_d$ and $\beta_d$, we introduce a coupling neck and adjust its radius without changing $\omega_m$ and $\beta_m$. This is because the volume of coupling neck is not large enough to modify monopole response, and it is located at the pressure anti-node (thus velocity node) of the monopole mode [see Fig. 1(a-b) and Fig. 4(f)]. In other words, we can achieve the nearly *independent* control of monopole and dipole mode frequencies and the dissipation factors until $\omega_m \approx \omega_d$ and $\beta_m \approx \beta_d$. The followed key step is to adjust $\ell$ and repeat the above steps until $\alpha_m \approx \alpha_d$ (the final $\ell = 0.4L$). Here, the designed CHR structure is duality symmetric but the critical coupling conditions are not met (see its scatterings and effective properties in Supplementary Figure 3). The final step is to fill the porous foam into the two cavities of the CHR [see Fig. 4(b)]. In this way, the extracted effective properties will simultaneously increase their $\beta_m$ and $\beta_d$, until the critical coupling criterions are met [i.e., $|S_m(\omega_m)| = |S_d(\omega_d)| \approx 0$].

## The observation of exceptional absorption bandwidth

After applying the design strategy, we have numerically and experimentally verified the scattering features of the CHR structure. The solid shell of the sample was fabricated by 3D printing technology and experiment details are given in Methods. As shown in Fig. 4(c), it is exhibited that the bandwidth of the measured absorption of our CHR sample is exceptionally large. The observed absorption with a near-double octave bandwidth ($A \geq 0.5$ from 1300 Hz to 4900 Hz) outperforms the reported ultrasparse sound-absorbing metamaterials[50,51] with the same mechanism of hybrid dipole-monopole resonances. In a followed section, we will further quantify this performance advantage.

For the absorption mechanism validation, the theoretical results of the modified surrogate model employing the fitting parameters ($\alpha_m = \alpha_d = 1.6$, $\omega_m = \omega_d = 2\pi \times 2500$Hz) are depicted with black solid and dashed lines in Fig. 4, in good agreement with the simulation (colored solid lines) and experimental data (circles) from low frequency to the resonance frequency with an absorption peak. Beyond 2500 Hz, the corresponding absorption is slightly lower than the measured one. The slight deviation is due to the simplification assumptions in the fitting approach (See Methods for the details of the fitting algorithm for effective properties). Nevertheless, by comparing Fig. 4(g) and Fig. 4(h), the spectral profiles of the extracted properties agree well, especially below 2500 Hz, i.e., $\frac{K_0}{K_{\text{eff}}(\omega)} \approx \frac{\rho_{\text{eff}}(\omega)}{\rho_0}$. The approximate compliance of duality symmetry can be evidenced by the low level of $|R(\omega)|$ in Fig. 4(e), and the aligned scattering amplitudes, i.e., $|S_m(\omega)| \approx |S_d(\omega)|$ as per Fig. 4(d). We recall that $|S_m(\omega)| \approx |S_d(\omega)|$ is one of the desirable conditions for extending absorption integral from the generalized causality constraint in Eq. (3). To evaluate how far our simulated absorption integral is from the generalized bound $\Gamma$, we introduce the absorption integral defined by

$$\gamma(\omega) = \int_0^{\omega} -\ln\big(1 - A(\omega)\big)\frac{d\omega}{\omega^2}. \tag{12}$$

Based on Eq. (12), we adopt $\lim_{\omega\to\infty} \gamma(\omega)/\Gamma$, as a measure of absorption optimality. The determination of $\Gamma$ is based on the extracted effective properties based on simulated scattering data at the low frequency limit. For our numerical evaluation in Fig. 4(f), we only integrate up to the frequency of 6000Hz, which is high enough to get the saturated value (48.9%Γ). Based on the modified surrogate model, we can also analytically determine absorption integral provided that the duality symmetry and critical coupling condition are satisfied. The absorption contributed by higher order terms is ignorable. Otherwise, the calculated $\gamma$ via simulation data (considering higher-order contribution) will deviate from the lowest-order theoretical value $\Gamma_1$. However, this is not true because we have $\gamma = 48.9\%\Gamma$, which is very close to $\Gamma_1 = 47.5\%\Gamma$. It is important to emphasize that this approximation is applicable solely to our specific case and does not yield a universal conclusion for other resonator structures. The closed-form derivation of $\Gamma_1$ is available in Supplementary Note 2. Therefore, the non-dispersive approximation of $\delta_m$ ($\delta_d$) is reasonable because from the low frequency limit to $\omega_m$ ($\omega_d$), the absorption contribution of the lowest-order monopole (dipole) resonance is dominant because of the weighting function $1/\omega^2$ in Eq. (12).

Because the high-order modes in a single CHR structure are difficult to tame towards ideal, perfect duality symmetry, we adopt an approach to circumvent this limitation to further broaden the absorption band. In Fig. 5, we further scale the geometry of the CHR structure to adjust the absorbing frequency. Four scaled CHR units are selected and integrated in series (see the details of tuning and integrating CHRs in Methods and Supplementary Figure 4, Figure 5, and Figure 6). As shown in Fig. 5(b), to inherit the duality symmetric design, each CHR has its different perforated rate of the narrowed channel to match the varying cavity volume [i.e., $\frac{K_0}{K_{\text{eff}}(\omega)} \approx \frac{\rho_{\text{eff}}(\omega)}{\rho_0}$ according to Eq. (11)]. Again, the final sample of integrated CHRs is filled with sufficient porous foam again to add enough dissipation and maintain critical coupling conditions.

In this way, in Fig. 5(d), we observe that the integrated CHRs demonstrate a broadband absorption (>50%) spanning from 300 Hz to 6000 Hz (measured average absorption: 86.1%). Our comparative experiment reveals that a standard foam liner, with the same dissipative volume and the total length, exhibits a lower average absorption of 61.4% within the same band. Furthermore, the calculated $\lim_{\omega\to\infty} \gamma(\omega)/\Gamma$ for the integrated CHRs (42.2%Γ) shows more than *double* enhancement than that of the foam liner (18.5%Γ), as shown in Fig. 5(e).

### A Figure of Merit (FOM) for two-port absorbers

The previous measure, $\lim_{\omega\to\infty} \gamma(\omega)/\Gamma$, while accurate and effective, is not compatible to the experimental data, because they are difficult to collect at all frequencies, especially for low frequency data. We conclude that our work is already the state of the art, with the aid of a Figure of Merit (FOM) we developed for two-port absorbers:

$$\text{FOM} = \sum_{n=1}^{N-1} -\ln\big(1 - A(\omega_n)\big)\frac{\omega_n - \omega_{n+1}}{\omega_n^2 \tilde{\Gamma}}\,, \tag{13}$$

where $A(\omega_n)$ is the absorption coefficient at the discretized frequency $\omega_n$ ($N$ data points), and the estimated geometry-dependent bound $\tilde{\Gamma} = \frac{\pi L}{2c_0}\left(\frac{K_0}{\tilde{K}_{\text{eff}}(0)} + \frac{\tilde{\rho}_{\text{eff}}(0)}{\rho_0}\right)$, based on the same principle of Eq. (11). It is likely that $\tilde{\Gamma}$ is an underestimated value because the contribution of end correction is omitted, as we mentioned. To quantify the resulting bias, we can compare the calculated $\Gamma$ of single CHR and integrated CHRs: the respective accurate values are $\Gamma = 6.1 \times 10^{-4}$[s] and $6.1 \times 10^{-4}$[s], while the FOM-based values are $\Gamma = 4.6 \times 10^{-4}$[s] and $4.0 \times 10^{-3}$[s]. This observed bias is both expected and justifiable, as near-field interactions contribute to end correction effects. Furthermore, rather than integrating over the entire frequency band, the simplified FOM places greater emphasis on high absorption and low-frequency bands (greater than 50%), resulting in higher scores for broadband absorbers. The detailed formulas and geometry inputs of $\tilde{\Gamma}$ are available in Supplementary Table 1. To be consistent for all evaluation cases, we consistently use the data in the frequency range defined with the threshold absorption of 50% [$A(\omega_n) \geq 0.5$ for arbitrary $n$].

By investigating the static values of $\frac{\tilde{\rho}_{\text{eff}}(0)}{\rho_0}$ and $\frac{K_0}{\tilde{K}_{\text{eff}}(0)}$, we classify our work, the foam liner, and others[25-30]. As shown in Fig. 6(a), the proposed structures meet the static duality symmetry (the dashed line). The foam liner and Ref.[30], mainly use volume-controlled compressibility to support sound absorption, while Refs.[25-29] take more advantage of the narrowed channel-controlled density to boost their performance. Those works[28,30] far away from the static duality symmetry line are doomed to obtain a low FOM score [see Fig. 6(b)]. However, our works (single CHR and integrated CHRs) with dispersion design of *broadband* duality symmetry as well as critical coupling conditions, rank the highest FOM. While our single CHR supports a single pair of monopole-dipole resonator, its FOM is higher than those with multiple resonant units, which has a comparable[26] or much lower bandwidth[25,27]. The integrated CHRs and the foam liner outperform all the others with the largest bandwidth, indicated by the width of the data bars, but the latter has a lower average absorption value of 61.3%. Even if Ref.[30] has a higher average absorption, the price to pay, is to use bulky resonators, as indicated by its large $\frac{K_0}{\tilde{K}_{\text{eff}}(0)}$. For a similar reason, Ref.[28] gains a considerably good balance in high absorption and large bandwidth, but its $\frac{\tilde{\rho}_{\text{eff}}(0)}{\rho_0}$ is very large, indicating a very low ventilation efficiency and leading to a low FOM. By contrast, Gao et al.[29] adopts an ultrasparse geometry with low $\frac{K_0}{\tilde{K}_{\text{eff}}(0)}$ and $\frac{\tilde{\rho}_{\text{eff}}(0)}{\rho_0}$, but its total length $L$ is very large, which unfortunately leads to a large $\tilde{\Gamma}$ as well. We have listed the data of Fig. 6 for FOM comparison in Supplementary Table 1. Many recently reported high-performance and innovative metamaterial absorbers[57,58] may also be re-evaluated and optimized using our figure of merit (FOM) evaluation framework. We foresee a promising future for acousto-mechanical metamaterials that transcends the limitations of conventional materials.

## Discussion

In summary, even our proposed structures are still some distance away from the upper limit of absorption ($\sim 1-\Gamma_1/\Gamma$), because we currently do not follow the simultaneous critical coupling and duality symmetry of high-order modes. Remarkably, the compliance of lowest-order monopole-dipole resonances has revealed the absorption potential not available in the previous works. The strategy of overlapping resonances to enhance wave-matter interaction is a well-established principle, leading to phenomena such as super-absorption[40] and super-scattering[41]. Our work builds on this principle by applying it to the specific case of acoustic monopole-dipole degeneracy. Moreover, this also provides ample room for optimization for future work, especially considering the development of artificial intelligence in helping to regulate the scattering spectrum of metamaterials[59,60]. For example, the impedance network could be effectively established with the multi-objective and coupling-driven optimization, while considering the additional requirements on the mechanical strength and lightweight design[61-63]. By using the example of ventilative acoustic metamaterials, we have put an end to the misleading situation where the 2-port scatterings are interpreted based on the one-port causality constraints[25-27]. We show that duality symmetry offers the unprecedented advantage of broadband anti-reflection and absorption effect, although dispersive resonators typically lead to strong scatterings. Our time-domain simulation further demonstrated this excellent feature (see Supplementary Movie).

For an ideal duality-symmetric system, the proposed generalized bound in Eq. (4) reduces to $\pi L Z_0/K_{\text{eff}}(0)$ or $\pi L \rho_{\text{eff}}(0)/Z_0$, which is a half of the original bounds of Eq. (1) or Eq. (2) (assuming identical sample thickness via $D \to L$). This explains why absorption is harder for an absorber without backing[20]: to achieve the same absorption performance, at least double thickness, or more generally speaking, the bound $\Gamma$, should be used to compensate the lowered bound; the common violation of broadband duality symmetry further hides the untapped absorption potential, as evidenced by Fig. 6. Based on the renewed understanding of the generalized bound, we realized that a handy FOM must be established to measure the performance of a two-port passive absorber, thus verifying the untapped absorption provided by the approximate compliance of duality symmetry.

The focus of our research is on one-dimensional two-port systems or their equivalent scattering problems characterized by reduced dimensions (see the comparison of waveguide and free space scattering in Supplementary Figure 7). In contrast, the more intricate landscape of higher-order scatterings in higher dimensions may present challenges that the straightforward pairing and cancellation between a monopole and a dipole—enforced by the duality symmetry—may not adequately address. Nevertheless, we foresee that the exploration of scattering causality and duality symmetry will provide physical insights for MIMO systems[64], potentially revealing additional constraints and symmetries[65-67].

This also presents substantial opportunities for real applications that are not limited to acoustic devices. The same principle also applies to electromagnetic wave transmission and solid vibration conduction. Moreover, some emerging advanced metamaterials[23,68] that break the passive, linear, time-invariant assumptions could follow a clearer direction for spectral optimization and a more reliable constraint for future benchmarking.

## Methods

### The simulation details and geometrical parameters

Simulation was carried out using the COMSOL Multiphysics commercial software, specifically employing the finite element method (FEM) facilitated through the 'Pressure Acoustics, Frequency Domain' module. The simulation setup included monitoring of incident and transmitted waves via two ports located at the boundaries of the model. The lossy components of the coupled Helmholtz resonator (CHR) structures, specifically the resonators' necks and the coupling neck between cavities, were calculated by using the narrow region model. In cases where the cavities were filled with a highly absorptive porous foam[69], the Johnson-Champoux-Allard (JCA) model was utilized, characterized by the following parameters (also adopted from Ref.[69]): porosity $\varphi = 0.94$, fluid resistance $R_\mathrm{f} = 32000\ [\mathrm{Pa}\cdot\mathrm{s/m^2}]$, tortuosity factor $\alpha_\infty = 1.06$, viscous characteristic length $L_v = 56[\mathrm{um}]$, thermal characteristic length $L_\mathrm{th} = 110[\mathrm{um}]$. These parameters were consistently applied across different simulations (including the foam liner in Fig. 5) to maintain a benchmark for comparison. For simulations in the time domain, the 'Pressure Acoustics, Time Explicit' interface was employed.

The geometrical specifications of the single CHR are as follows (see the marked dimensions in Supplementary Figure 2): diameter $D_0 = 1.5$ cm, length $L = 2.5$ cm, $\ell = 0.4L$, coupling channel diameter $d_2 = 0.625$ mm, width $w_1 = 1.5$ mm, thickness $t = 0.95$ mm, cavity height $H_c = 6.25$ mm, radius $R = 10.67$ mm, $\tau_1 = 3$ mm, $\tau_2 = 2$ mm. The perforated rate (or ventilation ratio) $\phi_0 = S/S_0 = (\pi H^2)/(\pi D^2) = 0.25$, where $S$ represents the cross-sectional area at the narrowest part of the main duct, and $S_0$ denotes the wave-front area or duct cross-sectional area. The thickness of other partition walls is standardized at $\Delta_w = 1\mathrm{mm}$. A $1/32$ reduced model features a sector angle of $360°/32 = 11.25°$.

The parameter combination outlined is not exclusive; however, the design adheres to the principles necessary to approach the bound of the generalized causality constraint, focusing on critical coupling and broadband duality symmetry. For achieving critical coupling, the methodology involves integrating sufficient porous foam within the resonator cavities to ensure maximal absorption ($S_m$ and $S_d$ = 0) at the resonance frequency, as demonstrated by the comparison between Fig. 4 in the main text and Supplementary Figure 3. Conversely, samples devoid of filled porous material exhibit duality symmetry but experience under-damped losses, resulting in degraded absorption performance.

The scaling of the sample in Fig. 5 adheres to the following rules: $V_c \to \alpha V_c$, $L \to \alpha^{1/2} L$, $\phi_0 \to \alpha^{-1/2}\phi_0$, where $\alpha$ is the scaling factor and $V_c$ is the cavity volume occupied by porous foam, calculated as $V_c = L(\pi(H_c + \tau_1 + D)^2 - \pi(\tau_1 + D)^2) + (L - l - 2\Delta_w)(\pi(D + \tau_1)^2 - \pi(D\sqrt{\phi_0} +$ $\Delta_w + \tau_1)^2$). The dimensions of the HR necks and the coupling channels between the two cavities remain unchanged during scaling. For the reference sample illustrated in Fig. 4, $\alpha = 1$; for scaled samples depicted in Fig. 5, $\alpha$ values are 0.5, 1.5, 2.5, and 4. The total thickness of the integrated sample is calculated as $L = (\sqrt{0.5} + \sqrt{1.5} + \sqrt{2.5} + \sqrt{4})L_0 + 3\Delta_w = 14.1\text{cm}$.

For the acoustic-flow interaction simulations (see Supplementary Figure 8), the flow resistance was calculated using the 'Linearized Navier-Stokes, Frequency Domain' interface. The background pressure drop ($\Delta P = P_2 - P_1$) was determined by the difference between the output and input ports. The background flow velocity ($v_f$) was varied from -20 m/s to 20 m/s, where negative values indicate flow in the opposite direction to the propagation of the acoustic wave. The flow resistance of the integrated CHR structures was calculated as $R_f = \Delta P/(v_f L)$. The absorption spectra for five selected flow velocities were computed using the 'Compressible Potential Flow' interface, in conjunction with the 'Linearized Potential Flow, Frequency Domain' interface.

### The extraction and fitting of effective acoustic properties

The simulated reflection, and transmission coefficients [$R(\omega)$ and $T(\omega)$] were defined at the front and back surfaces of the sample, respectively. These coefficients facilitated the extraction of effective bulk modulus and effective density spectra ($\frac{K_0}{K_{\text{eff}}(\omega)}$ and $\frac{\rho_{\text{eff}}(\omega)}{\rho_0}$), utilizing the formula outlined in Ref.[56]. It is crucial to distinguish these coefficients from the S-parameters, which are directly defined at the surfaces of the input and output ports. The relationships are expressed as $R(\omega) = S_{11}e^{-2ik_0L_{wg}}$ and $T(\omega) = S_{21}e^{-2ik_0L_{wg}}$ [$L_{wg}$ is the distance between the sample surface to the input or output port, see Supplementary Figure 2]. For the extraction of effective properties in the experiments, the data were processed using the same algorithm as applied, except that $L_{wg}$ was substituted with the distance from the sample surface to the nearest microphone.

The extracted effective properties from the CHR structures are fitted with the Lorentz model with the locked broadband duality symmetry (i.e., $\alpha_{m/d} = \alpha_m = \alpha_d$, $\omega_{m/d} = \omega_m = \omega_d$, $\delta_{m/d} = \delta_m = \delta_d$). In addition, the dissipation factors ($\beta_{m/d} = \beta_m = \beta_d$) are fixed according to closed-form solutions from Supplementary Note 2. The target of our fitting is to minimize the cost function: $\sum_{n=1}^{N}\left|\frac{K_0}{K_{\text{eff}}(\omega_n)} - \frac{\alpha_m\omega_m^2}{\omega_m^2-\omega_n^2-i\beta_m\omega_n} - \delta_m\right|^2 + \left|\frac{\rho_{\text{eff}}(\omega_n)}{\rho_0} - \frac{\alpha_d\omega_d^2}{\omega_d^2-{\omega_n}^2-i\beta_d\omega_n} - \delta_d\right|^2$. The fitting results ($\alpha_{m/d}$, $\omega_{m/d}$, $\delta_{m/d}$) were obtained via the optimization toolbox (Levenberg–Marquardt algorithm for non-linear fitting) provided by the commercial software of MATLAB. We approximately treated $\delta_m$ and $\delta_d$ as non-dispersive constants (the fitted value $\delta_{m/d} \approx 1$), thus ignoring the higher-order contribution without losing the insight of causality-guided physics. The rationale for this assumption is that the absorption integral is mainly due to the contribution of low-frequency modes and that the absorption of a single monopole-dipole resonator gradually decreases at high frequencies, because treating $\delta_{m/d}$ as a constant without dispersion and a complex dispersion fitting model that does not conform to the duality symmetry ($\delta_m(\omega) \neq \delta_d(\omega)$) have the same result, with no evident contribution to absorption integral. So, finally, the corresponding fitted outcomes of the modified surrogate model are plotted in Fig. 4 with black solid and dashed lines.

### Sample fabrication and experimental characterization

The fabrication of the sample frame was facilitated by employing 3D printing technology, specifically Stereolithography (SLA) using WeNext 8228 Resin. This method provided the precision required to accurately resolve features such as narrow channels and necks with dimensions smaller than 0.4 mm. The resulting structures exhibited sufficient rigidity to be considered as hard boundaries in the context of acoustic fields, where they are associated with thermoviscous boundary layers. In the experiment, a custom-built impedance tube with a circular cross-section and a diameter of 3cm was employed. The experimental apparatus comprised BSWA microphones (model MPA416), NI sound cards (model CompactDAQ9263), and an NI acquisition card (model CompactDAQ9234). Harmonic sine waves were generated utilizing a HIVI speaker (model M3N) in conjunction with a YAMAHA power amplifier (model PX3). The acoustic properties of the samples were characterized using a standard four-microphone technique within the impedance tube measurements. Prior to each series of experiments, microphone mismatch calibration was meticulously performed in accordance with established standards (ASTM E2611-09). To reduce noise interference due to leakage, Plasticine was applied to seal all potential connection gaps. Furthermore, the cut-off frequency of the impedance tube was determined to be 6700 Hz, which is marginally above the upper frequency limit of our absorption measurements (6000 Hz). Data below 300 Hz were not adopted due to the challenges and inaccuracies associated with low-frequency measurements. To compare the experiment-based data with simulation predictions, an offset value $\Delta\gamma_{\mathrm{simu}}$ from simulation data was utilized. For comparisons illustrated in Fig. 4(f) and Fig. 5(e), the adopted formula (data presented by circles) was that $\gamma(\omega) = \int_{2\pi\times 300}^{\omega} -\ln\left(1 - A_{\mathrm{exp}}(\omega)\right)\frac{d\omega}{\omega^2} + \Delta\gamma_{\mathrm{simu}}$, where $A_{\mathrm{simu}}$, $A_{\mathrm{exp}}$ denote absorption spectra from simulation and experiments respectively and $\Delta\gamma_{\mathrm{simu}} = \int_{0}^{2\pi\times 300} -\ln\left(1 - A_{\mathrm{simu}}(\omega)\right)\frac{d\omega}{\omega^2}$.

### Data availability

All the data are provided with this paper and Supplementary Information and Supplementary File. We have added Source Data of the geometry of the acoustic resonators (see Supplementary Data). Together with the details in Methods, this represents the minimum dataset that is sufficient to reproduce all simulation and experimental results.

## Acknowledgements

This work was supported by Jockey Club Trust STEM Lab of Scalable and Sustainable Photonic Manufacturing (GSP181). S. Q. thanks Doris Zimmern HKU-Cambridge Hughes Hall Fellowships and Seed Fund for Basic Research for New Staff from HKU-URC (No. 103035008). S. Q and N. X. F. acknowledge the financial support from RGC Strategic Topics Grant (STG3/E-704/23-N) and ITC-ITF project (ITP/064/23AP). I. D. A. acknowledges the support from the Royal Society, London, for Industry Fellowship. E. D. and N. X. F. thank the startup funding from MILES in HKU-SIRI. S. Q. thank Dr. Ruo-Yang Zhang for inspiring discussion on duality symmetry and Dr. Yang Meng for acoustic-flow interaction.

## Author contributions

S.Q., P.S., convinced the idea of the duality-induced generalized causality constraint. S.Q., M.Y., P.S., I.D.A., N.X.F. performed the theoretical model. S.Q., E. D. performed the acoustic FEM simulations. S.Q., S. H., H. Y. L., performed the data postprocessing and results analysis. S.Q., S.L. performed the material fabrication and acoustic experiments. S.Q. wrote the manuscript. S.Q., M.Y., I.D.A., N.X.F., P.S. reviewed, and edited the manuscript.

## Competing interests

The authors declare no competing interests.

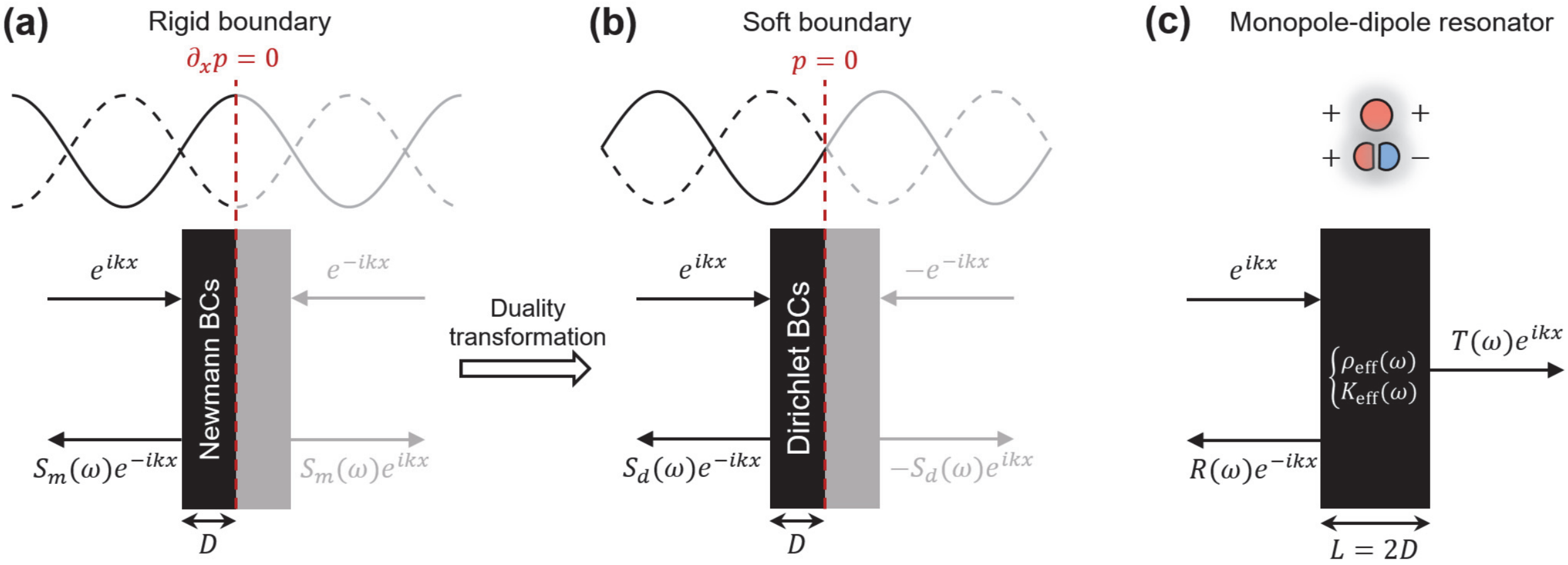

**Figure 1. Wave scattering setups. (a)** The monopole scattering with symmetrical pressure fields. **(b)** The dipole scattering with anti-symmetrical pressure fields. The sound fields in either half of the spaces in (a) and (b) can be applied to the causality constraints as per Eq. (1) and Eq. (2) (equivalent to the 1-port setup). **(c)** The effective medium description of a monopole-dipole resonator, with generalized 2-port setup allowing both reflection and transmission.

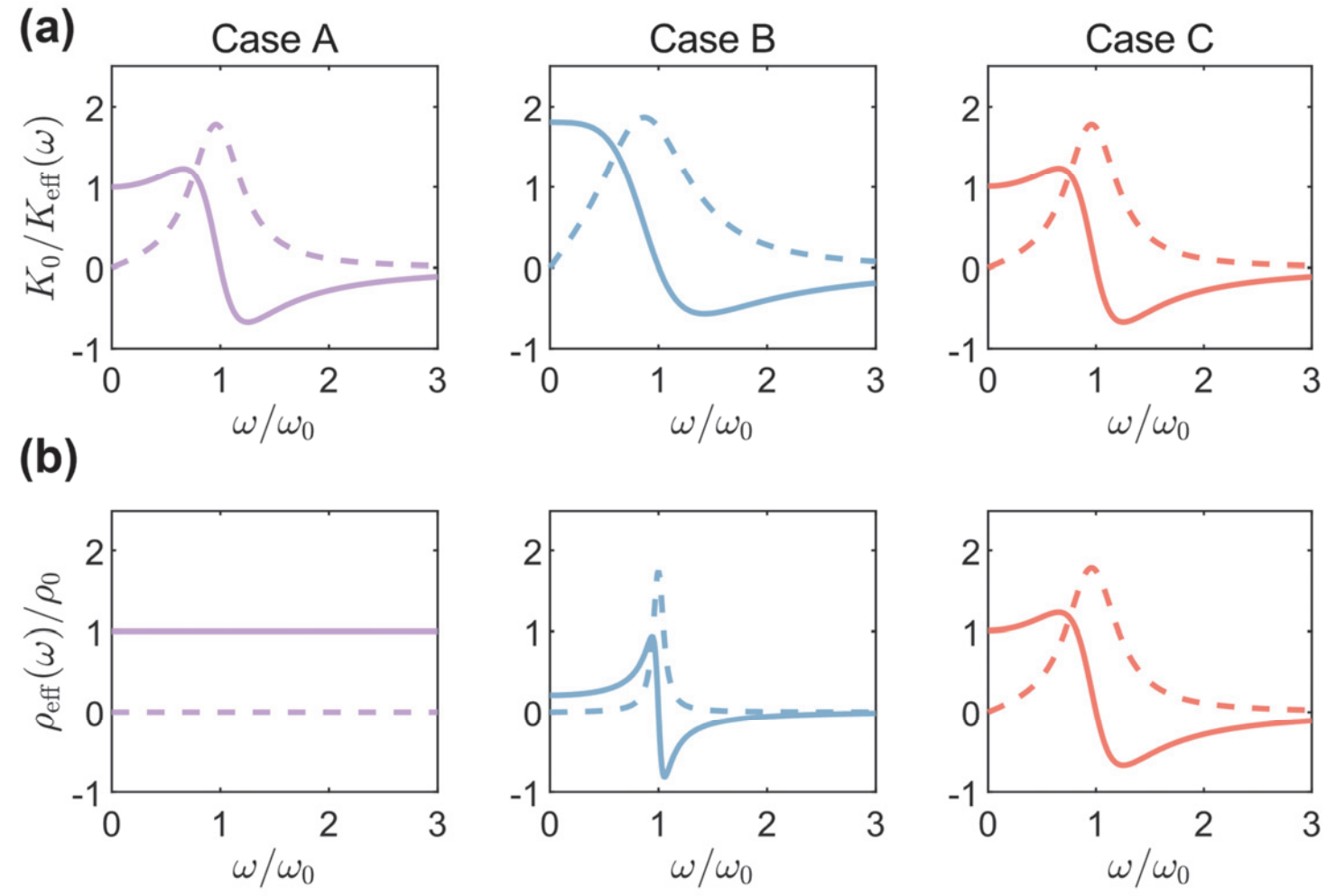

**Figure 2. The dispersion profiles of Lorentzian model. (a)** Effective compressibility and **(b)** density of Case A (a single-mode monopole resonator), Case B (a narrowband duality-symmetric monopole-dipole resonator), and Case C (a broadband duality-symmetric monopole-dipole resonator).

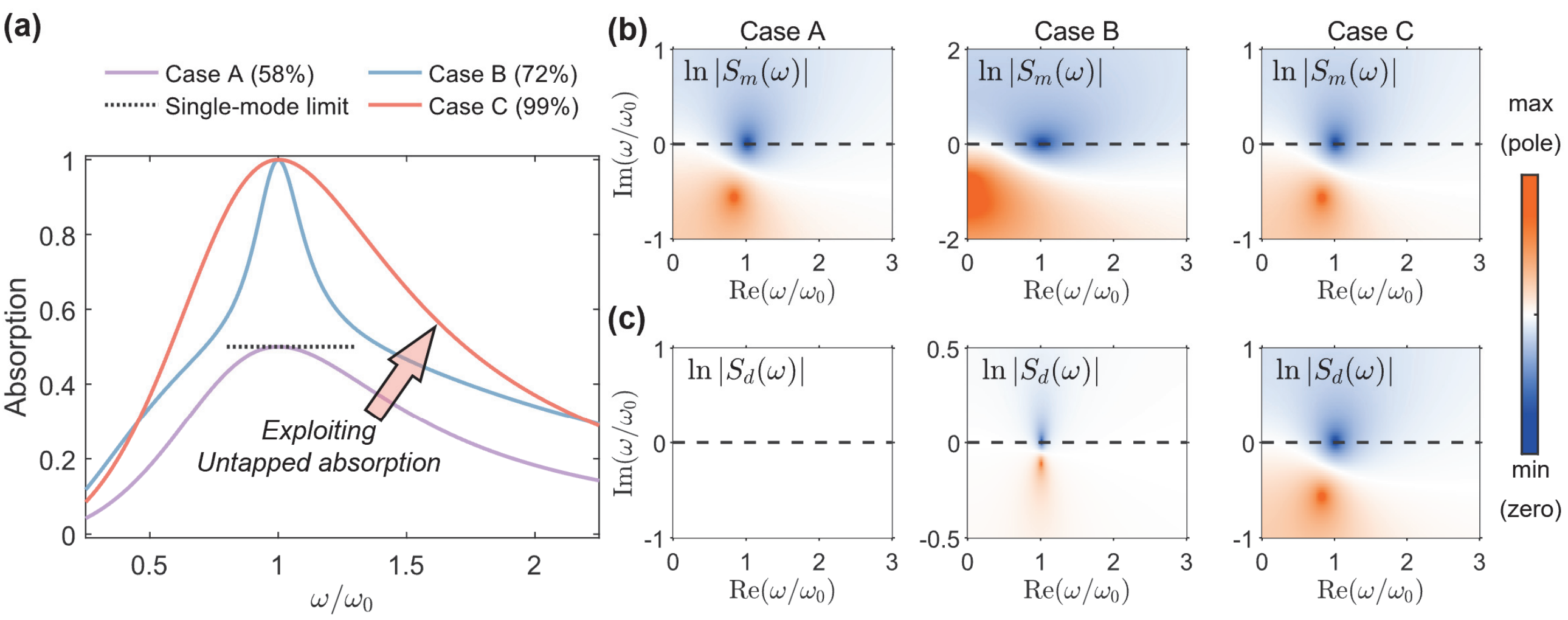

**Figure 4. Absorption spectra and the zero-pole profiles of two-port scatterings. (a)** The corresponding absorption spectra. The legend shows the percentage between the full band integration of Eq. (12) ($\omega \to \infty$) and $\Gamma$. **(b-c)** The complex frequency plane showing the distribution of $\ln|S_m(\omega)|$ and $\ln|S_d(\omega)|$. All the zeros are forced to be critically coupled on the real axis by using Eq. (7). In the first panel of (e), the constant effective density causes the non-existence of dipole resonance mode with blank data of the zero and pole.

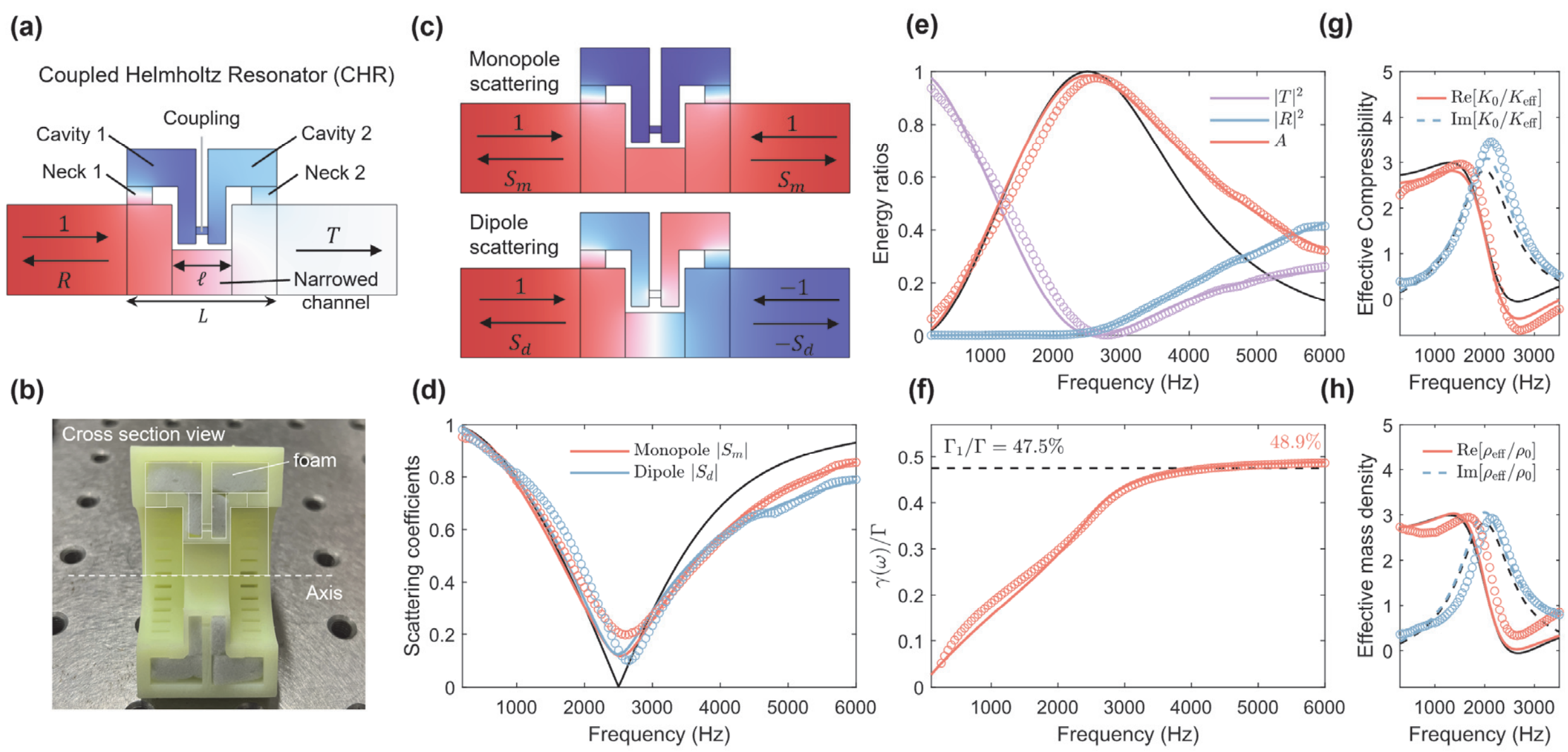

**Figure 3. The experimental realization of duality-symmetric sample. (a)** The schematic of coupled Helmholtz resonators (CHR). The color indicates the total pressure mode at maximal absorption frequency. The thickness $L = 2.5$cm. **(b)** The fabricated sample (half model) filled with porous foam. **(c)** The monopole and dipole modes at 2500 Hz (pressure field). **(d)** The corresponding scattering strength spectra of CHR. **(e)** The scattering spectra of CHR (transmission, reflection, and absorption). **(f)** The ratio between absorption integral $\gamma(\omega)$ and $\Gamma$. **(g)** The extracted effective density and **(h)** effective compressibility. The black lines represent the results by the modified surrogate model, the colored lines are simulation data, and the circles are measured data by the impedance tube measurement.

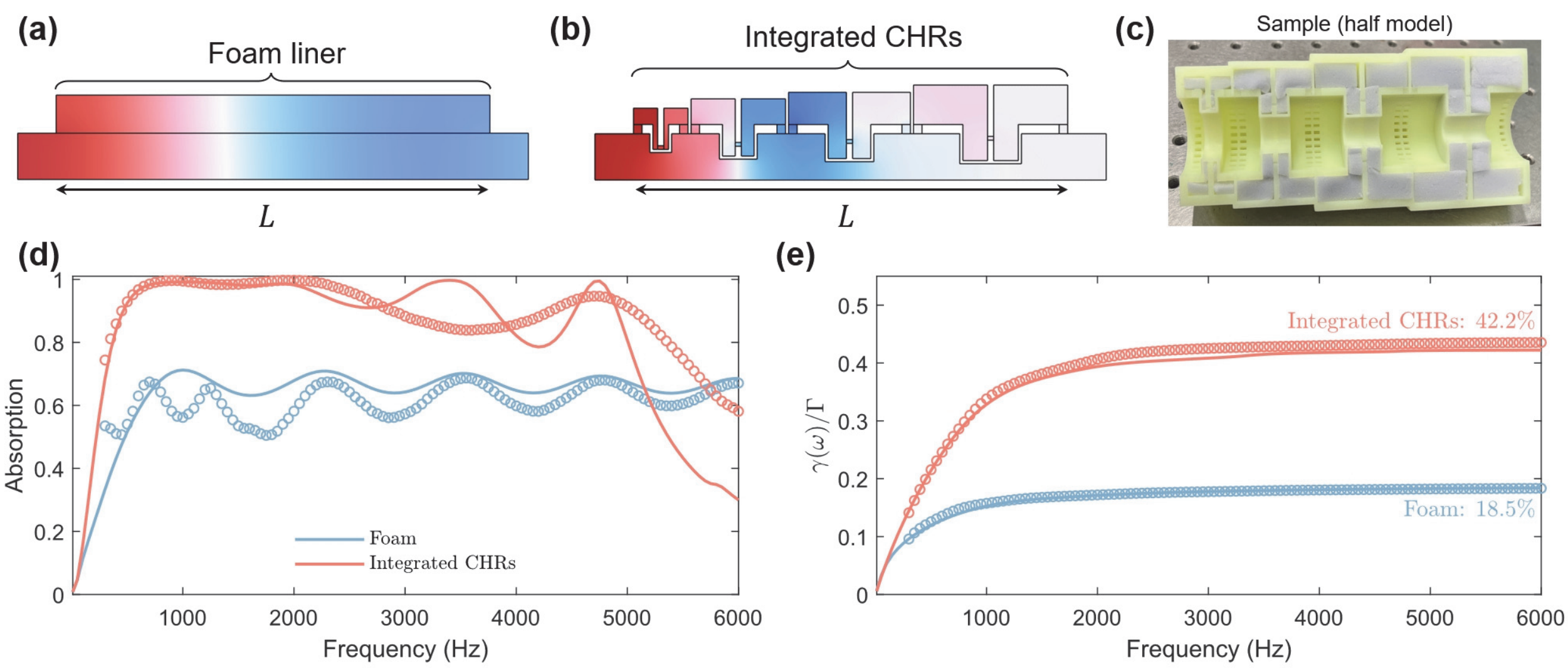

**Figure 5. The characterization of a foam liner and the integrated CHRs. (a-b)** The pressure fields of foam liner and integrated CHRs at 1000Hz (simulated). The total thickness is the same, i.e., $L = 14.1\text{cm}$. **(c)** the cross section of fabricated integrated CHRs (half model for the illustration). **(d)** Absorption performance comparison between traditional porous foam and our integrated CHRs. **(e)** The integrals of the absorption spectra over the bound $\Gamma$. The solid lines are simulation data while the circles are from the experimentally collected ones.

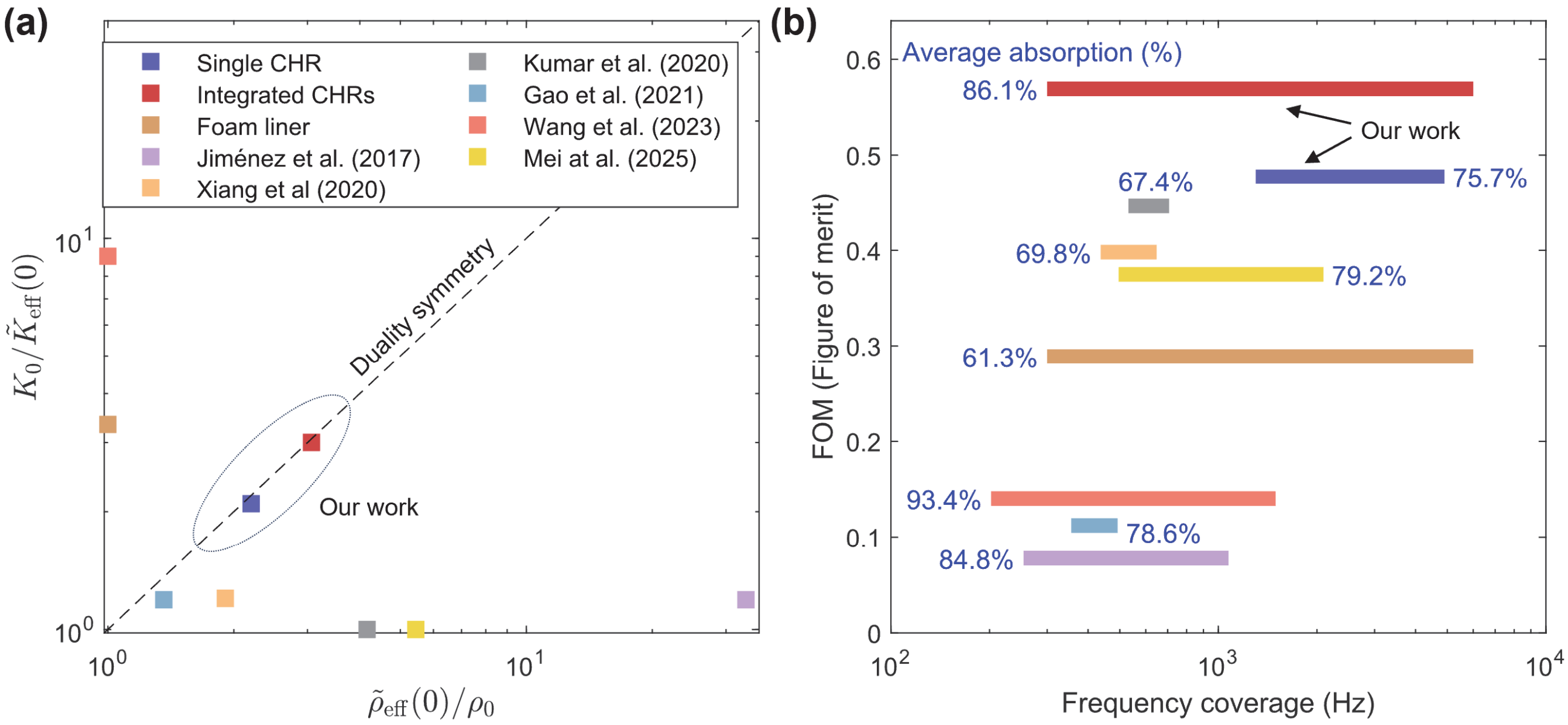

**Figure 6. The performance evaluation of the two-port absorbers. (a)** The absorbers (our single CHR, integrated CHR, foam liner and other competitive works), classified by their static effective density and compressibility. **(b)** The further comparison of their absorption performance based on the proposed FOM in Eq. (13).

# Supplementary Information

# Duality symmetry reveals untapped absorption potential from a generalized causality constraint

Sichao Qu, [1,2] Min Yang, [3,*] Sibo Huang, [4] Shuohan Liu, [1] Erqian Dong, [1] Helios Y. Li, [1]
Ping Sheng, [2,†] I. David Abrahams, [2,‡] and Nicholas X. Fang[1,5,§]

[1]*Department of Mechanical Engineering, The University of Hong Kong, Pokfulam Road, Hong Kong, China*
[2]*Department of Applied Mathematics and Theoretical Physics (DAMTP), University of Cambridge, Wilberforce Road, Cambridge CB3 0WA, UK*
[3]*Acoustic Metamaterials Group, Data Technology Hub, No. 5 Chun Cheong Street, Hong Kong, China*
[4]*Department of Electrical Engineering, City University of Hong Kong, Tat Chee Avenue, Kowloon, Hong Kong, China*
[5]*Materials Innovation Institute for Life Sciences and Energy (MILES), HKU-SIRI, Shenzhen, China*

Correspondence to: [*]min@metacoust.com; [†]sheng@ust.hk; [‡]ida20@cam.ac.uk; [§]nicxfang@hku.hk

**Supplementary Note 1. The duality transformation for causality constraints**

We construct that $\mathcal{D}(\theta) = SR(\theta)S^{-1}$, where the 2D rotation matrix $R(\theta) = \begin{pmatrix} \cos\theta & -\sin\theta \\ \sin\theta & \cos\theta \end{pmatrix}$, and $S = \begin{pmatrix} 1 & 0 \\ 0 & e^{\frac{i\pi}{2}} \end{pmatrix} = \begin{pmatrix} 1 & 0 \\ 0 & i \end{pmatrix}$ (one typical type of the logic gate matrix for $\frac{\pi}{2}$ phase conversion). It should be noted that in electromagnetic case, the duality transformation matrix coincides $R(\theta)$ [1]. We interpret that $S$ depicts the difference of field orientation between longitudinal waves (acoustics) and transverse waves (electromagnetism). In this way, the continuous duality transformation is characterized by the matrix:

$$\mathcal{D}(\theta) = \begin{pmatrix} \cos\theta & i\sin\theta \\ i\sin\theta & \cos\theta \end{pmatrix}, \tag{S1}$$

where $\theta$ is a tunable parameter. We adopt the general duality transformation on the pressure and velocity fields as follows (in the main text, typical values $\theta = 0, \frac{\pi}{2}$ are adopted):

$$\begin{bmatrix} p_\theta \\ v_\theta \end{bmatrix} = \mathcal{D}(\theta) \begin{bmatrix} p \\ v \end{bmatrix} = \begin{pmatrix} \cos\theta & iZ_0 \sin\theta \\ i\sin\theta / Z_0 & \cos\theta \end{pmatrix} \begin{bmatrix} p \\ v \end{bmatrix}, \tag{S2}$$

where the appearance of $Z_0$ is due to the unit conversion between $p$ and $v$. In Eq. (S2), the dimensionless $\mathcal{D}(\theta)$ can be restored by treating $Z_0 \to 1$. Consequently, the specific impedance on the surface transitions to:

$$Z_\theta = \frac{p_\theta}{v_\theta} = \frac{\cos\theta\, p + iZ_0 \sin\theta\, v}{i\sin\theta\, p/Z_0 + \cos\theta\, v}. \tag{S3}$$

Applying a duality transformation to 1D acoustic equations results in:

$$\partial_x \begin{pmatrix} p_\theta \\ v_\theta \end{pmatrix} = i\omega \left[ \mathcal{D}(\theta) \begin{pmatrix} 0 & \rho \\ K^{-1} & 0 \end{pmatrix} \mathcal{D}^{-1}(\theta) \right] \begin{pmatrix} p_\theta \\ v_\theta \end{pmatrix}, \tag{S4}$$

which simplifies to:

$$\partial_x \begin{pmatrix} p_\theta \\ v_\theta \end{pmatrix} = i\omega \begin{pmatrix} i\dfrac{\sin\theta\cos\theta}{c_0}\left(\dfrac{K_0}{K} - \dfrac{\rho}{\rho_0}\right) & \rho_0 \left(\sin^2\theta \dfrac{K_0}{K} + \cos^2\theta \dfrac{\rho}{\rho_0}\right) \\ \dfrac{1}{K_0}\left(\sin^2\theta \dfrac{K_0}{K} + \cos^2\theta \dfrac{\rho}{\rho_0}\right) & -i\dfrac{\sin\theta\cos\theta}{c_0}\left(\dfrac{K_0}{K} - \dfrac{\rho}{\rho_0}\right) \end{pmatrix} \begin{pmatrix} p_\theta \\ v_\theta \end{pmatrix}. \tag{S5}$$

Here, the matrix of material properties includes Willis coupling (diagonal terms) because of the duality transformation on acoustic fields and material property matrix.

Addressing the monopole causality case initially, the integral inequality can be restated as (refer to Ref. [2]):

$$\int_0^{\infty} \frac{-\ln(|S_m(\omega)|^2)}{\omega^2}\, d\omega = \frac{2\pi D}{c_0} \frac{K_0}{K_{\text{eff}}(0)} - \pi \sum_n \frac{\text{Im}(\omega_n)}{|\omega_n|^2}, \qquad (S6)$$

where $\omega_n$ represents the zeros of $\ln(|S_m|)$ in the upper complex frequency plane. Zeros located in the lower half-plane are excluded from consideration. Here, $\text{Im}(\omega_n) > 0$ serves as the basis for the inequality form of the causality constraint. For a lossless system, where the amplitude of $S_m(\omega)$ is unity, the causality constraint simplifies to:

$$\Gamma_m = \frac{2\pi D}{c_0} \frac{K_0}{K_{\text{eff}}(0)} = \sum_n \Gamma_m^{(n)} = \pi \sum_n \frac{\text{Im}(\omega_n)}{|\omega_n|^2}, \qquad (S7)$$

with $\Gamma_m^{(n)} = \pi \frac{\text{Im}(\omega_n)}{|\omega_n|^2}$. Thus, the integral forms of the causality constraints for monopole and dipole scattering provide crucial insights into the physical properties of the system under study.

Decomposing the bound $\Gamma_m$ into contributions from intrinsic resonances within the systems is feasible. Upon integrating loss into the practical system, the same conceptual framework is applicable, although variations will occur in the ratios of $\text{Im}(\omega_n)/|\omega_n|^2$. Certain terms may disappear because of shifts into the lower half of the complex frequency plane, which are not included in the set of $\omega_n$. The critical coupling condition for the $n$-th order resonance $\omega_n$ is expressed as $\text{Im}(\omega_n) = 0$, indicating that the zero of $\omega_n$ lies precisely on the real frequency axis where $S_m(\omega_n) = 0$.

In the context of dipole scattering causality, the analysis proceeds similarly, yielding the equation:

$$\int_0^{\infty} -\ln|S_d(\omega)|^2 \frac{d\omega}{\omega^2} = \frac{2\pi D}{c_0} \frac{\rho_{\text{eff}}(0)}{\rho_0} - \pi \sum_n \frac{\text{Im}(\omega_n')}{|\omega_n'|^2}, \qquad (S8)$$

where the set of $\omega_n'$ here pertains to the zeros of $\ln|S_d|$ in the upper complex frequency plane. Consequently, we define with lossless case that,

$$\Gamma_d = \frac{2\pi D}{c_0} \frac{\rho_{\text{eff}}(0)}{\rho_0} = \sum_n \Gamma_d^{(n)} = \pi \sum_n \frac{\text{Im}(\omega_n')}{|\omega_n'|^2}, \qquad (S9)$$

where $\Gamma_d^{(n)} = \pi \frac{\text{Im}(\omega_n')}{|\omega_n'|^2}$. The generalized causality constraint is reformulated as:

$$\begin{aligned} \int_0^{\infty} -\ln\big(1 - A(\omega)\big) \frac{d\omega}{\omega^2} &\leq \frac{1}{2} \int_0^{\infty} -[\ln|S_m(\omega)| + \ln|S_d(\omega)|] \frac{d\omega}{\omega^2} \\ &= \Gamma - \sum_n \left(\Gamma_m^{(n)} + \Gamma_d^{(n)}\right), \end{aligned} \qquad (S10)$$

where the sum $\sum_n (\Gamma_m^{(n)} + \Gamma_d^{(n)})$ will vanish if the zeros of both monopole and dipole resonances are critically coupled or over-damped (located on or below the real axis). Therefore, the bound $\Gamma$ is defined as:

$$\Gamma = \Gamma_m + \Gamma_d = \frac{2\pi D}{c_0} \left[ \frac{\rho_{\text{eff}}(0)}{\rho_0} + \frac{K_0}{K_{\text{eff}}(0)} \right]. \qquad (S11)$$

In our experimental design, only first-order monopole and dipole resonances are considered, where critical coupling of first-order resonances is achieved if $\Gamma_1 = \Gamma_m^{(1)} + \Gamma_d^{(1)} = 0$. Thus, the target bound is adjusted as:

$$\int_0^{\infty} -\log\big(1 - A(\omega)\big)\frac{d\omega}{\omega^2} \leq \Gamma_1 = \Gamma - \Gamma_{\infty}, \tag{S12}$$

where $\Gamma_{\infty} = \sum_{n\geq 2} \left(\Gamma_m^{(n)} + \Gamma_d^{(n)}\right)$, which can be ascertained through analytical solutions in a subsequently modified surrogate model considering higher-order resonances.

In systems exhibiting duality symmetry, the relationship $\rho_{\text{eff}}(\omega)/\rho_0 = K_0/K_{\text{eff}}(\omega)$ and $S_m(\omega) = -S_d(\omega)$ allows us to deduce that $\Gamma_m^{(n)} = \Gamma_d^{(n)}$ for any arbitrary $n$. Consequently, the higher-order bound is given by

$$\Gamma_{\infty} = \sum_{n\geq 2} \left[\Gamma_m^{(n)} + \Gamma_d^{(n)}\right] = \frac{2\pi \mathrm{Im}(\tilde{\omega})}{|\tilde{\omega}|^2} = \frac{2\pi}{\mathrm{Im}(\tilde{\omega})}, \tag{S13}$$

where $\tilde{\omega} = ip$ $(p > 0)$. The generalized causality constraint is expressed as

$$\int_0^{\infty} \left[-\ln\big(1 - A(\omega)\big)\right]\frac{d\omega}{\omega^2} \approx \Gamma - \Gamma_{\infty} = \frac{2\pi D}{c_0}\left[\frac{\rho_{\text{eff}}(0)}{\rho_0} + \frac{K_0}{K_{\text{eff}}(0)}\right] - \frac{2\pi}{p}, \tag{S14}$$

achieving equality only under the condition of duality symmetry. The first-order bound $\Gamma_1 = \Gamma - \Gamma_{\infty}$ (numerical value) closely aligns with the absorption integral of our sample (experiment value), as outlined in Fig. 4(f).

**Supplementary Note 2. The surrogate model with higher-order resonances**

The surrogate model incorporates Lorentzian formulations to represent the dispersion of material properties while ensuring compliance with the Kramer-Kronig relations. This is expressed mathematically as:

$$\begin{cases} \dfrac{K_0}{K_{\text{eff}}(\omega)} = \dfrac{\alpha_m \omega_m^2}{\omega_m^2 - \omega^2 - i\beta_m \omega} + \displaystyle\sum_{i>1} \dfrac{\alpha_{m,i}\omega_{m,i}^2}{\omega_{m,i}^2 - \omega^2 - i\beta_{m,i}\omega} \\ \dfrac{\rho_{\text{eff}}(\omega)}{\rho_0} = \dfrac{\alpha_d \omega_d^2}{\omega_d^2 - \omega^2 - i\beta_d \omega} + \displaystyle\sum_{i>1} \dfrac{\alpha_{d,i}\omega_{d,i}^2}{\omega_{d,i}^2 - \omega^2 - i\beta_{d,i}\omega} \end{cases}. \tag{S15}$$

The higher-order terms $\delta_m$ and $\delta_d$ are considered constant when $\omega_{m,i} \gg \omega_m$, and $\omega_{d,i} \gg \omega_d$ (for $i \geq 2$). Hence, this formulation can be simplified as Eq. (13). This approximation effectively captures the profiles and values of absorption integrals of the effective properties between our sample and the modified surrogate model. In the results of Fig. 3, these higher-order terms are assumed to be negligible, modeling an ideal scenario with only first-order monopole or dipole resonances.

For monopole scattering scenario, as per Fig. 1(a), the coefficient

$$S_m = \frac{Z_s - Z_0}{Z_s + Z_0}, \tag{S16}$$

where the surface impedance $Z_s = iZ_{\text{eff}}\cot(k_{\text{eff}}D)$ according to the impedance transfer theory [3]. By taking the *subwavelength approximation*, we have

$$S_m = \frac{iZ_{\text{eff}}\cot(k_{\text{eff}}D) - Z_0}{iZ_{\text{eff}}\cot(k_{\text{eff}}D) + Z_0} \approx \frac{c_0 + i\dfrac{K_0}{K_{\text{eff}}(\omega)}(\omega D)}{c_0 - i\dfrac{K_0}{K_{\text{eff}}(\omega)}(\omega D)}. \tag{S17}$$

For dipole scattering scenario, as per Fig. 1(b), the coefficient can be derived similarly

$$S_d = \frac{-iZ_{\text{eff}}\tan(k_{\text{eff}}D) + Z_0}{iZ_{\text{eff}}\tan(k_{\text{eff}}D) - Z_0} \approx -\frac{c_0 + i\dfrac{\rho_{\text{eff}}(\omega)}{\rho_0}(\omega D)}{c_0 - i\dfrac{\rho_{\text{eff}}(\omega)}{\rho_0}(\omega D)}. \tag{S18}$$

Under the same subwavelength approximation, it follows that $S_d \to -S_m$, fulfilling the requirements of the duality transformation. Comparison of Eq. (S17) and Eq. (S18) substantiates that the duality symmetry necessitates:

$$Z_{\text{eff}}(\omega) = Z_0 \Rightarrow \frac{K_0}{K_{\text{eff}}(\omega)} = \frac{\rho_{\text{eff}}(\omega)}{\rho_0} \tag{S19}$$

Note: this relationship is valid for $S_m(\omega) = -S_d(\omega)$ with arbitrary frequency even outside the subwavelength approximation.

Now focusing on the effective bulk modulus term, the insertion of the first line of Eq. (13) into Eq. (S17) yields the analytical form for monopole scattering:

$$S_m = \frac{c_0 + i\left(\dfrac{\alpha_m \omega_m^2}{\omega_m^2 - \omega^2 - i\beta_m \omega} + \delta_m\right)(\omega D)}{c_0 - i\left(\dfrac{\alpha_m \omega_m^2}{\omega_m^2 - \omega^2 - i\beta_m \omega} + \delta_m\right)(\omega D)}, \tag{S20}$$

which simplifies to a Padé approximant of order [3/3] in the variable of $z$ $(z = -i\omega)$:

$$S_m = \frac{P(z)}{Q(z)}, \tag{S21}$$

where the numerator $P(z) = a_3 z^3 + a_2 z^2 + a_1 z + a_0$ and the denominator $Q(z) = -a_3 z^3 + b_2 z^2 + b_1 z + a_0$. Here, $a_3 = -\delta_m D$, $a_2 = c_0 - \beta_m \delta_m D$, $a_1 = c_0 \beta_m - (\alpha_m + \delta_m)\omega_m^2 D$, $a_0 = c_0 \omega_m^2$, $b_2 = c_0 + \beta_m \delta_m D$, $b_1 = c_0 \beta_m + (\alpha_m + \delta_m)\omega_m^2 D$. One of the three roots of the cubic equation (i.e., $P(\tilde{z}) = 0$) must be real (the corresponding $\tilde{\omega}$ should be purely imaginary), and the other two roots should be conjugate pair. Therefore, $P(z)$ can be reformulated as

$$P(z) = a_3 (z - p)(z - q + ir)(z - q - ir), \tag{S22}$$

where $p$, $q$ and $r$ should be real. By equating this expression with the original form, we have the following relations: $a_2 = -(p + 2q)a_3$, $a_1 = (q^2 + r^2 + 2pq)a_3$, $a_0 = -p(q^2 + r^2)a_3$.

If critical coupling condition applies, $q = 0$, and $r = \sqrt{a_1/a_3} = \sqrt{\frac{(c_0\beta_m - (\alpha_m + \delta_m)\omega_m^2 D)}{-\delta_m D}}$ $\left(\beta_m < \frac{(\alpha_m + \delta_m)\omega_m^2 D}{c_0}\right)$.

The perfect impedance matching will be achieved at a real frequency $\omega = r$ (assuming $r > 0$). Moreover, the critical condition in terms of Routh-Hurwitz criterion [4] can be derived: $a_1 a_2 = a_0 a_3$, which yields a quadratic equation in terms of $\beta_m$:

$$\beta_m^2 - \left[\frac{(\alpha_m + \delta_m)\omega_m^2 D}{c_0} + \frac{c_0}{\delta_m D}\right]\beta_m + \frac{\alpha_m \omega_m^2}{\delta_m} = 0. \tag{S23}$$

The roots are given by

$$\begin{cases} \beta_c^+ = \dfrac{1}{2}\left[\left(\dfrac{(\alpha_m + \delta_m)\omega_m^2 D}{c_0} + \dfrac{c_0}{\delta_m D}\right) + \sqrt{\Delta}\right] \\ \beta_c^- = \dfrac{1}{2}\left[\left(\dfrac{(\alpha_m + \delta_m)\omega_m^2 D}{c_0} + \dfrac{c_0}{\delta_m D}\right) - \sqrt{\Delta}\right] \end{cases}, \tag{S24}$$

where $\Delta = \left(\frac{(\alpha_m + \delta_m)\omega_m^2 D}{c_0}\right)^2 + \frac{2(-\alpha_m + \delta_m)\omega_m^2}{\delta_m} + \left(\frac{c_0}{\delta_m D}\right)^2$. We select the smaller value $\beta_c^-$ as analytical condition of critical coupling, instead of $\beta_c^+$, because $\beta_c^+$ will lead to complex-valued $r$. We have visualized the decomposition of the causality constraint bound in Supplementary Figure 1 with and without critical coupling (proper loss).

For the dipole scattering scenario, the critical coupled loss value $\beta_d$, as per Fig. 1(b), can similarly be derived by substituting $\omega_m \to \omega_d$, $\delta_m \to \delta_d$, $\alpha_m \to \alpha_d$. The analytical $\beta_c^-$, for the modified surrogate model with non-zero $\delta_m, \delta_d$, is pivotal for all numerical results in Fig. 4 (black lines), while for zero $\delta_m, \delta_d$, we use Eq. (10) instead.

For dual symmetric case, we define that $\delta_{m/d} = \delta_m = \delta_d = \tilde{\delta}$, $\alpha_{m/d} = \alpha_m = \alpha_d = \tilde{\alpha}$, $\beta_{m/d} = \beta_m = \beta_d = \beta_c^-$ according to Eq. (S19), we have

$$p = -\frac{a_2}{a_3} = \frac{c_0}{\tilde{\delta}D} - \beta_c^-, \tag{S25}$$

which determines the first-order bound:

$$\Gamma_1 = \Gamma - \Gamma_\infty = 2\pi \left[\frac{L(\tilde{\delta} + \tilde{\alpha})}{c_0} - \left(\frac{c_0}{\tilde{\delta}D} - \beta_c^-\right)^{-1}\right], \tag{S26}$$

where $\beta_c^-$ adopts the value in Eq. (S24). We used Eq. (S26), divided by $\Gamma$, to evaluate the level of the dashed horizontal line in Fig. 4(f).

**Supplementary Table 1.** The comparison of our work, the foam liner, and other competitive reported metamaterial absorbers [5-10] (all are with 2-port setups). This comparison elucidates a trade-off between the volume $V_{\mathrm{air}}$, ventilation rate $\phi_i$ and the absorption integral $\int_{\lambda_1}^{\lambda_2} -\ln(1-A)\,d\lambda$. The thickness $L$ of these materials is specified as the minimal distance required to geometrically contain the absorbing domains along the direction of wave propagation. The absorption band is specified by the frequency range $[f_1, f_2]$ or wavelength range $[\lambda_1, \lambda_2]$, where the absorption coefficient exceeds 50% (as a standard threshold). The evaluation formula $\frac{\rho_{\mathrm{eff}}}{\rho_0} \cong \sum \frac{\eta_i}{\phi_i}$ is the generalized version of the mass density in Eq. (11) of the main text, to consider the multi-layer perforated structures. All displayed absorption data are experimental from our measurements while others are extracted from the cited references via the software (i.e., WebPlotDigitizer [11], an open-source tool to extract numerical data from plots and graph images). The geometry details of Refs. [5-10] are directly available in their respective geometry description texts. The newly proposed $\mathrm{FOM} = \frac{\int_{\lambda_1}^{\lambda_2} -\ln(1-A)d\lambda}{L\left(\frac{\tilde{\rho}_{\mathrm{eff}}}{\rho_0}+\frac{K_0}{\tilde{K}_{\mathrm{eff}}}\right)} \approx \sum_{n=1}^{N-1} -\ln\left(1-A(\omega_n)\right)\frac{\omega_n-\omega_{n+1}}{\omega_n^2\tilde{\Gamma}}$ , where he estimated geometry-dependent bound $\tilde{\Gamma} = \frac{\pi L}{2c_0}\left(\frac{K_0}{\tilde{K}_{\mathrm{eff}}(0)} + \frac{\tilde{\rho}_{\mathrm{eff}}(0)}{\rho_0}\right)$.

| Reference | $[f_1, f_2]$ with $A > 0.5$ | $L$ | $\int_{\lambda_1}^{\lambda_2} -\ln(1-A)\,d\lambda$ | $\phi_i$ | $\eta_i$ | $\frac{\tilde{\rho}_{\mathrm{eff}}}{\rho_0} \cong \sum \frac{\eta_i}{\phi_i}$ | $\frac{K_0}{\tilde{K}_{\mathrm{eff}}} \cong \frac{V_{\mathrm{air}}}{S_0 L_{\mathrm{eff}}}$ |
|---|---|---|---|---|---|---|---|
| Our work (Single CHR) | 1300-4900 | 2.5cm | 0.51m | [1,0.25] | [0.6,0.4] | 2.2 | 2.1 |
| Our work (Integrated CHRs) | 300-6000 | 14.1cm | 4.81m | [1 0.35 0.2 0.16 0.13] | [0.6 0.05 0.09 0.12 0.15] | 3.0 | 3.0 |
| Foam liner | 300-6000 | 14.1cm | 1.74m | 1 | 1 | 1.0 | 3.34 |
| Ref. [5] | 254-1074 | 11.3cm | 3.0m | [0.16 0.16 0.15 0.14 0.13 0.13 0.13 0.01] | 1/8 for each | 33.5 | 1.19 |
| Ref. [6] | 437-648 | 7.5cm | 0.92m | 0.524 | 1 | 1.91 | 1.2 |
| Ref. [7] | 513-707 | 3cm | 0.68m | 0.24 | 1 | 4.2 | 1.18 |
| Ref. [8] | 358-490 | 47.2cm | 1.33m | 0.74 | 1 | 1.36 | 1.19 |
| Ref. [9] | 202-1500 | 37.2cm | 5.15m | 1 | 1 | 1.0 | 9 |
| Ref. [10] | 500-2078 | 5.3cm | 1.26m | [0.08 0.16 0.25 0.36 0.45] | 1/5 for each | 1.224 | 1.18 |

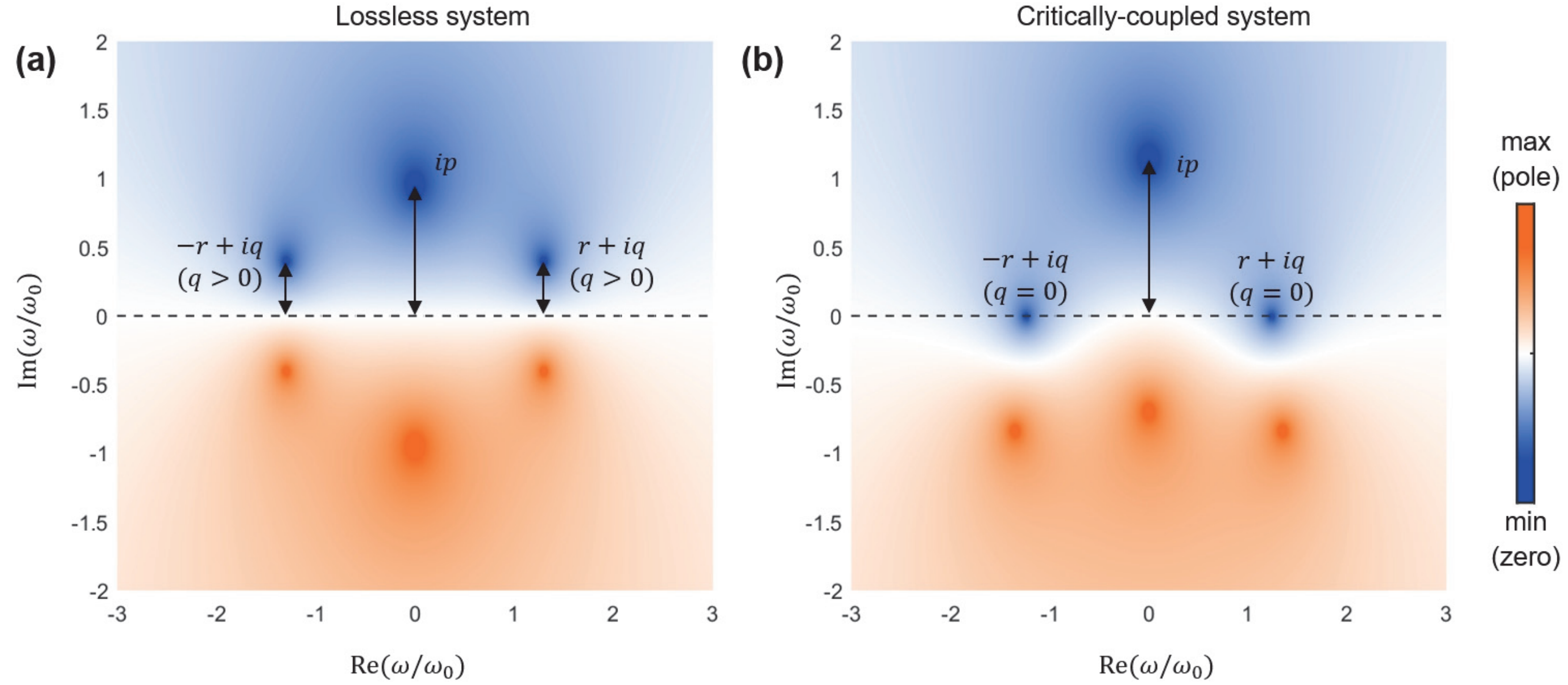

**Supplementary Figure 1. The complex frequency plane showing $\ln(|S_m(\omega)|)$ or $\ln(|S_d(\omega)|)$, with the colormap indicating maxima (poles) and minima (zeros). (a)** Represents a lossless system. **(b)** Illustrates a critically coupled system. The distance of the zeros to the real axis contributes to the decomposition of the total bound. The zero located on the imaginary axis corresponds to the solution $(z = p)$ obtained analytically from the roots of a cubic equation. For the numerical experiment, we utilized parameters from the modified surrogate mode.

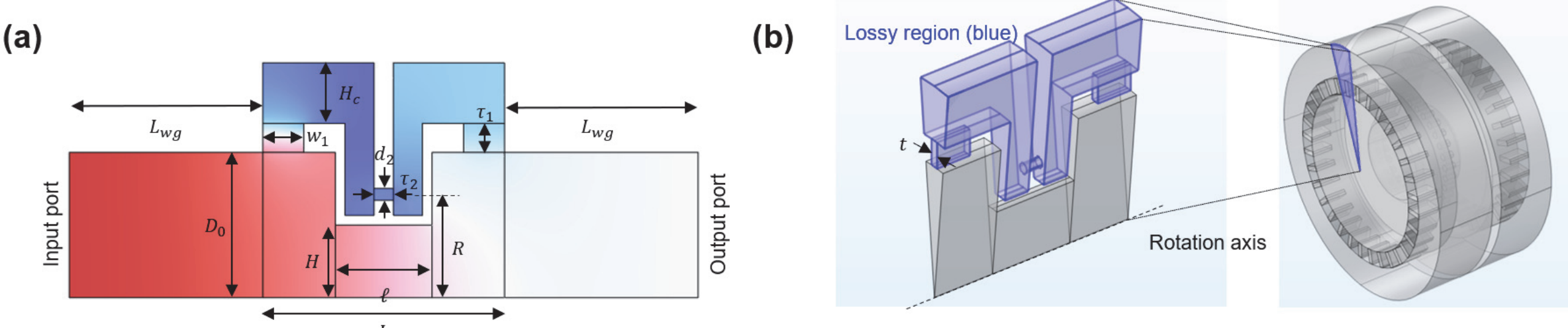

**Supplementary Figure 2. (a)** Geometrical parameters of the Coupled Helmholtz Resonators (CHR) geometry featuring input and output ports (left). Total pressure field in the transmission port is cancelled due to the opposite phases of monopole and dipole scatterings ($S_m \cong -S_d$). **(d)** The full model of air domains in a practical sample (right panel) and a reduced $1/32$ 3D model used for efficient simulation (left panel), with the blue regions indicating lossy components calculated by narrow region model. In the dual cavities, we used JCA model. Note: we modelled the main duct as lossless parts, because within the simulated band, the typical channel length $\gg \delta_v$ (thermoviscous boundary layer thickness).

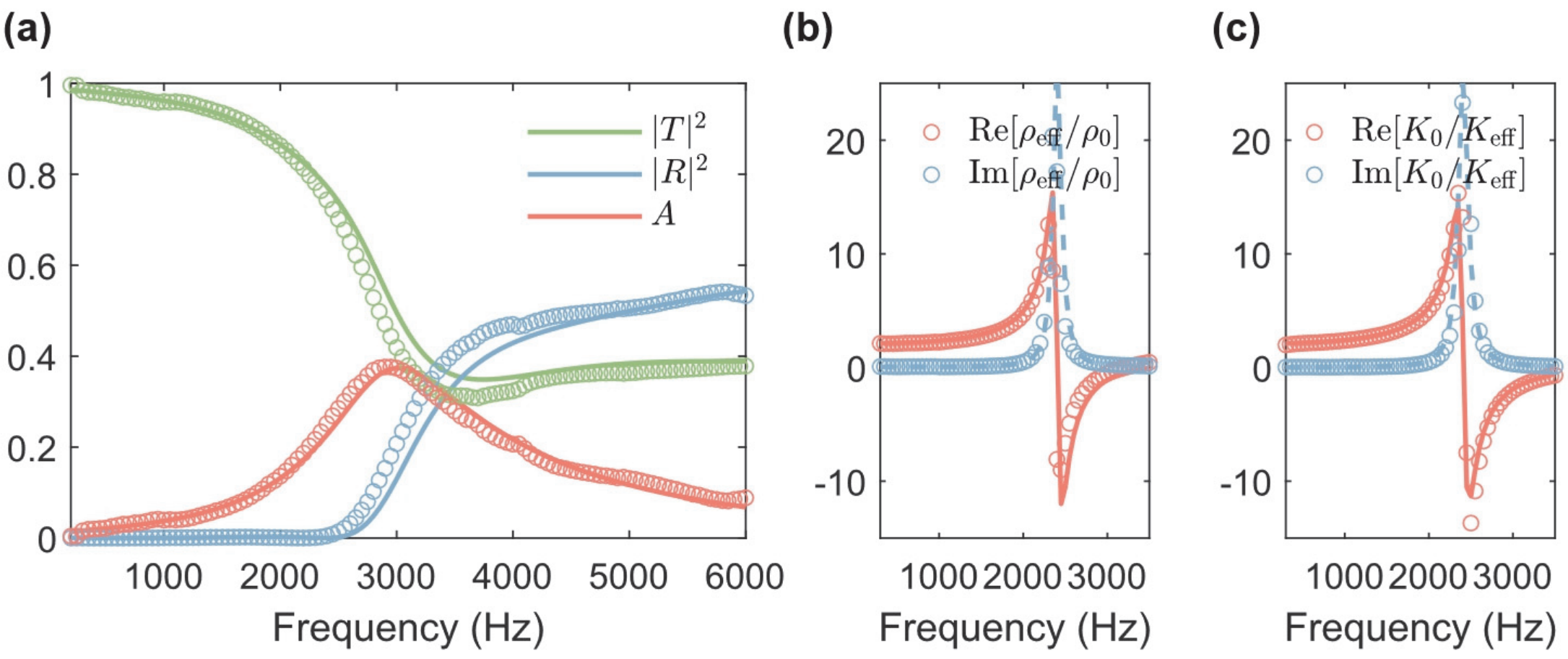

**Supplementary Figure 3. Coupled Helmholtz resonators (CHR) without filled porous foam in the cavities** (identical to the structures in Fig. 3). **(a)** Scattering spectra of CHR. Below the resonant frequency, the reflection is still extremely low, indicating the preservation of duality symmetry. **(b-c)** Effective density and effective compressibility profiles presented as solid lines for simulation results and circles for experimental data. The graphs illustrate the absence of critical coupling and highlight duality symmetry $[\rho_{\text{eff}}(\omega)/\rho_0 \cong K_0/K_{\text{eff}}(\omega)]$.

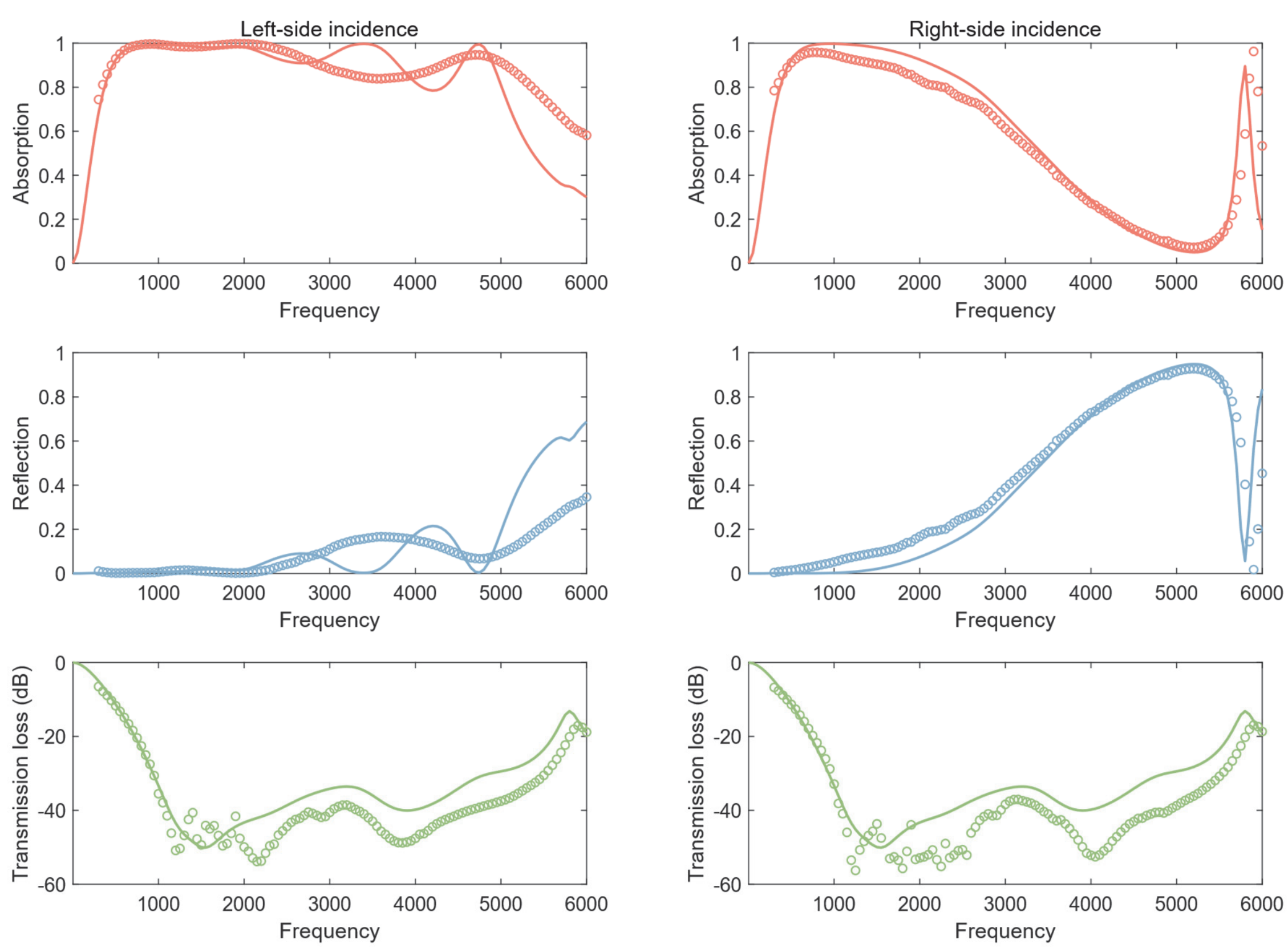

**Supplementary Figure 4.** The left panel denotes the scattering data of integrated CHR from left-side incidence (the same setup as the main text). The right panel represents the same with inversed excitation direction (right incidence). The absorption (the first row) exhibits asymmetric feature, due to the distinct reflection (the second row). The transmission data agree well for left and right incidence cases, due to the protection of reciprocity. The circles are measured data while the solid lines are extracted from simulation.

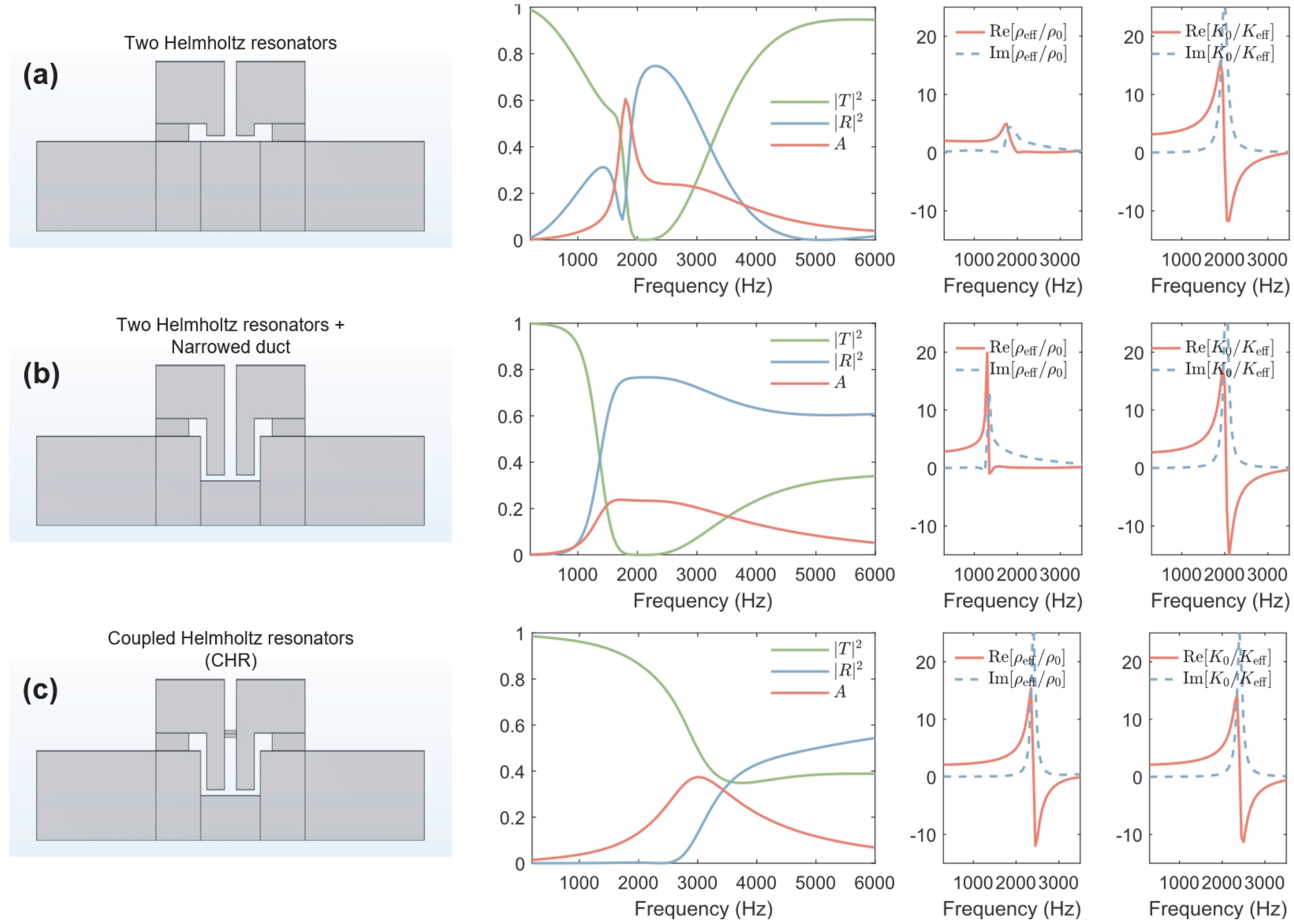

**Supplementary Figure 5. Design scheme of duality symmetry-protected sample. (a)** Standard double Helmholtz resonator mounted on the sidewall of a waveguide. **(b)** Identical resonators with unchanged cavity volume, featuring a narrowed channel in the main duct. **(c)** Coupled Helmholtz resonator with both a narrowed channel and a coupling neck, identical to the configuration in Supplementary Fig. 3. Note: In all cases, the cavities are not filled with foam.

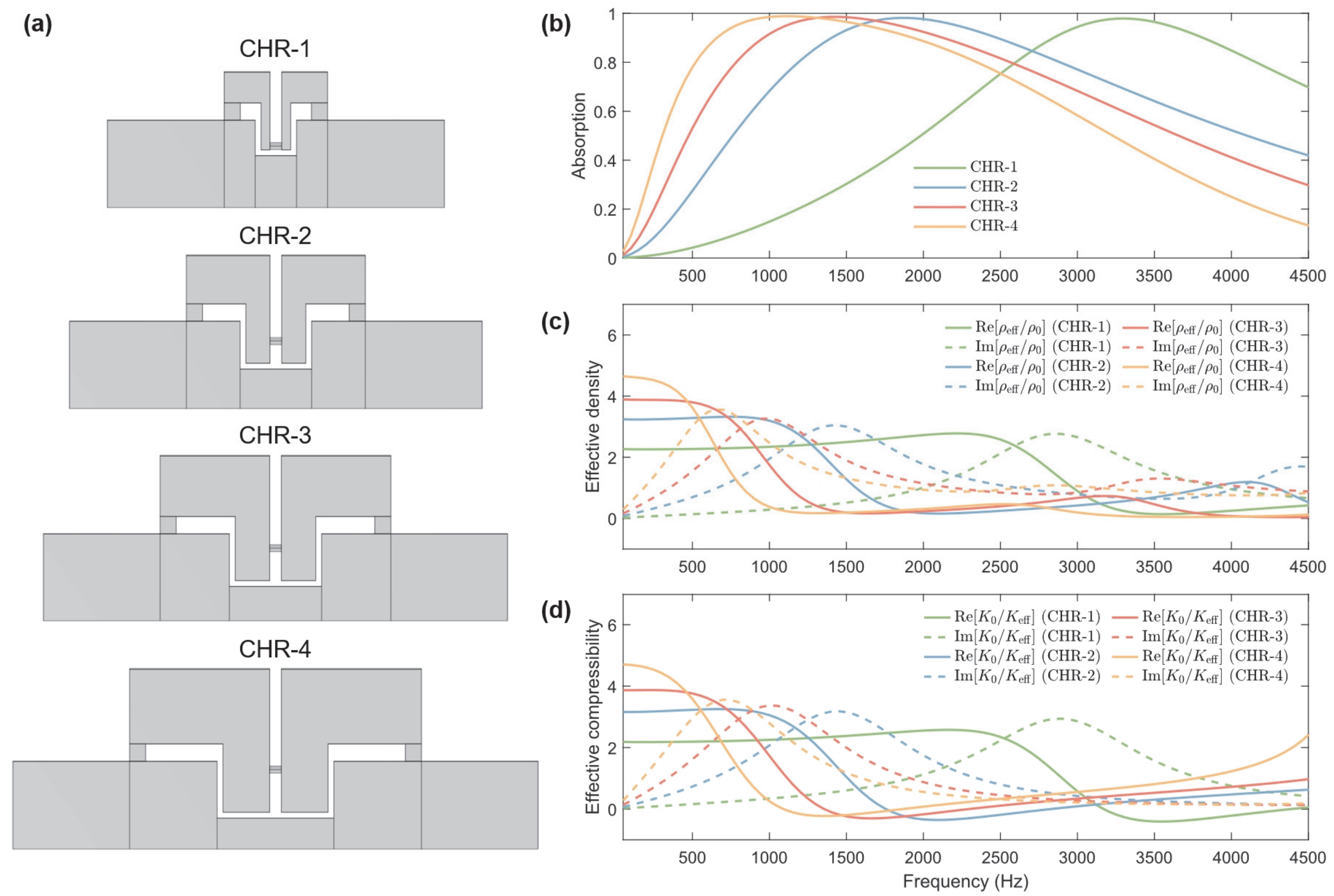

**Supplementary Figure 6. (a)** The geometry of scaled CHRs with tunable resonances. **(b)** The absorption spectra of each CHR. **(c-d)** The duality symmetry protected effective density and compressibility of each CHR, showing the practical realization of broadband degeneracy with tunable monopole and dipole modes.

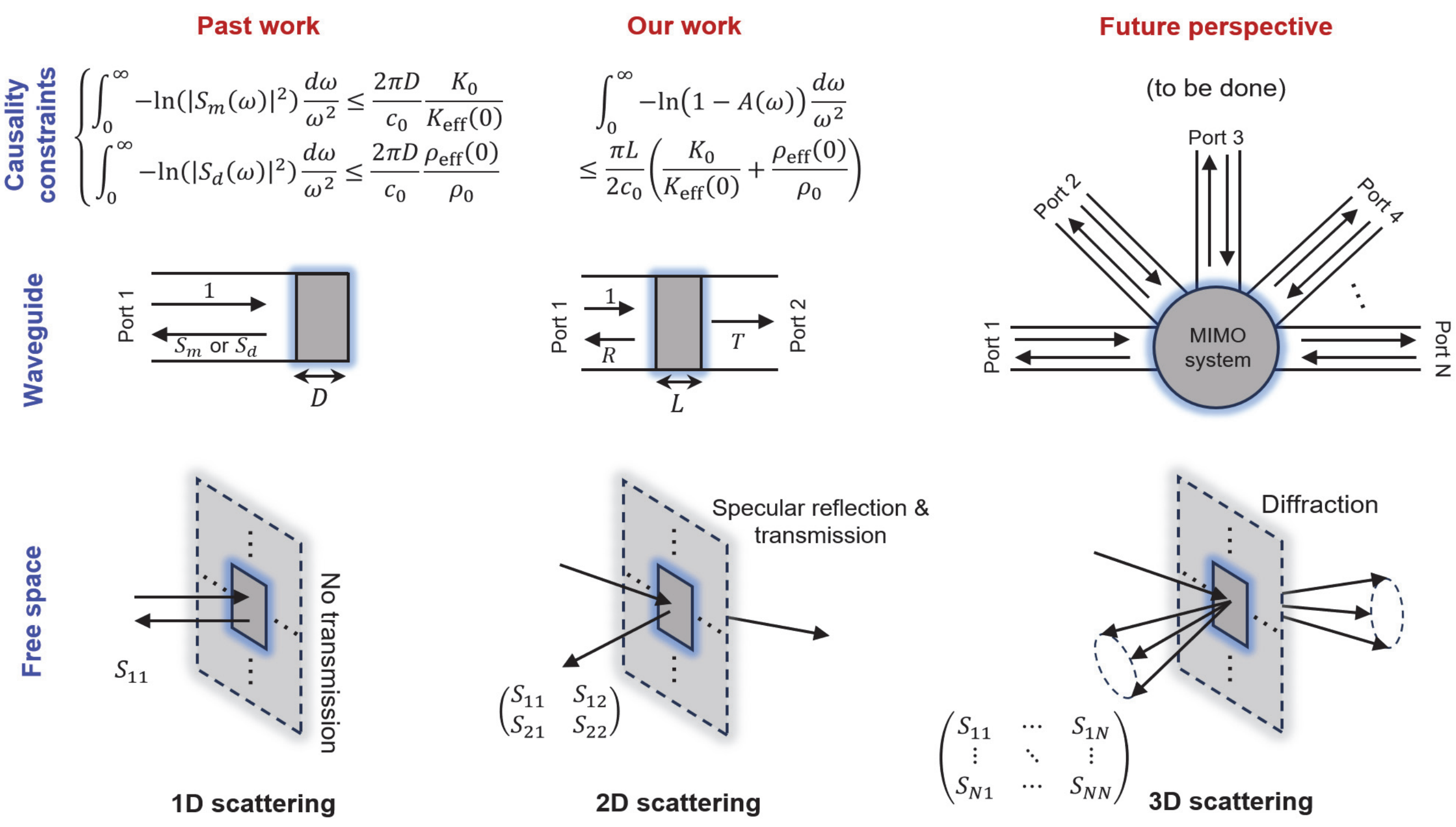

**Supplementary Figure 7. The scattering setups of previous research, our current focus, and the prospective causality-constraint generalization to higher-order dimensions and multi-port systems.** Notably, free-space scattering can be effectively reduced to waveguide transport of waves, as both share the same mathematical formulation of the scattering matrix, provided that the dimensions ($N \times N$) are identical.

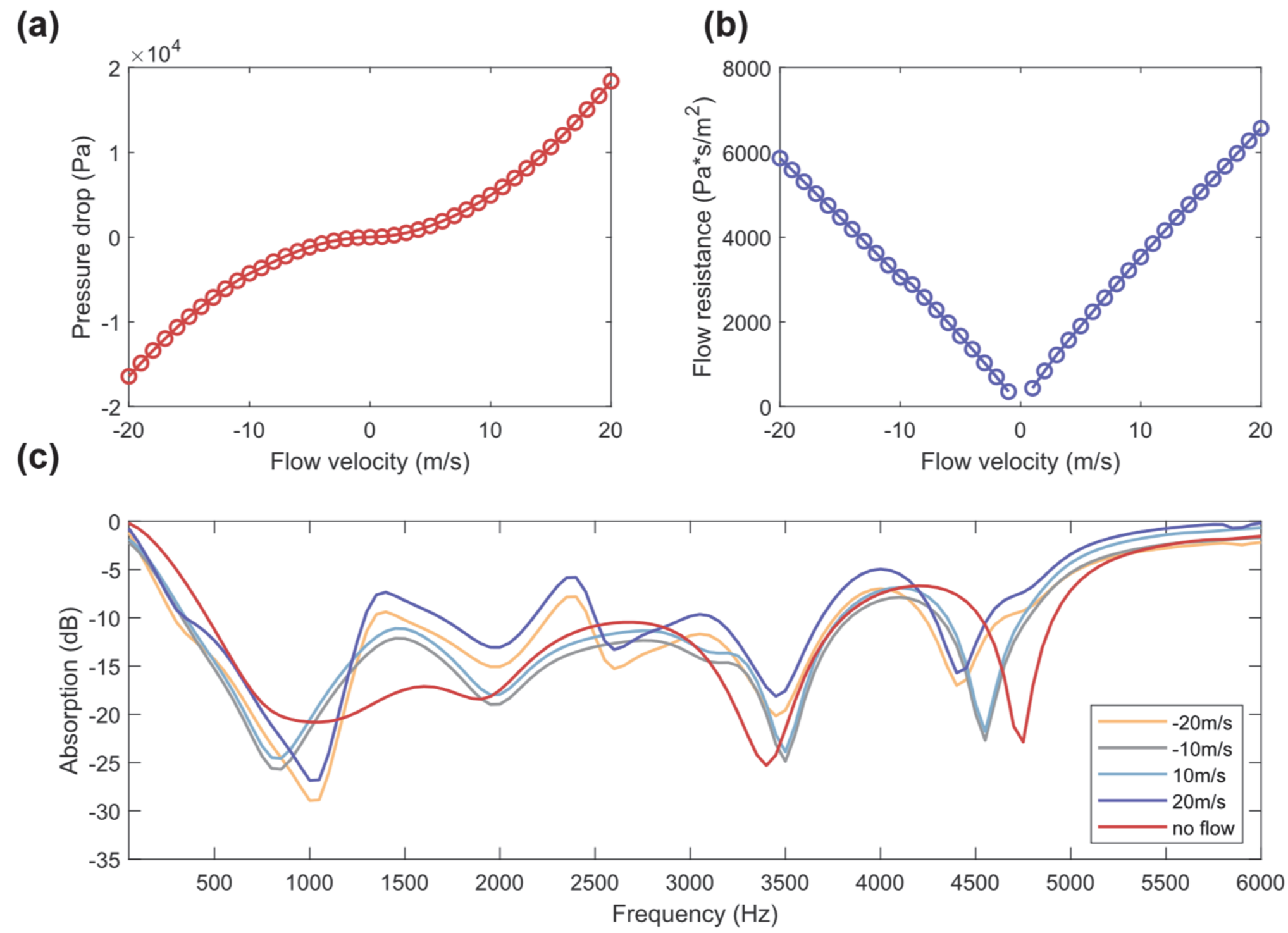

**Supplementary Figure 8. The interaction between the background flow and the acoustic absorption structures (integrated CHRs). (a)** The relationship between the pressure drop (Pa) and the background flow velocity (m/s). A positive flow velocity indicates that the flow is in the same direction as sound propagation, while a negative flow velocity indicates the flow is in the opposite direction. **(b)** The corresponding fluid resistance as a function of flow velocity. **(c)** The absorption spectra at selected flow velocities. Simulation details are provided in the Methods section of the main text.

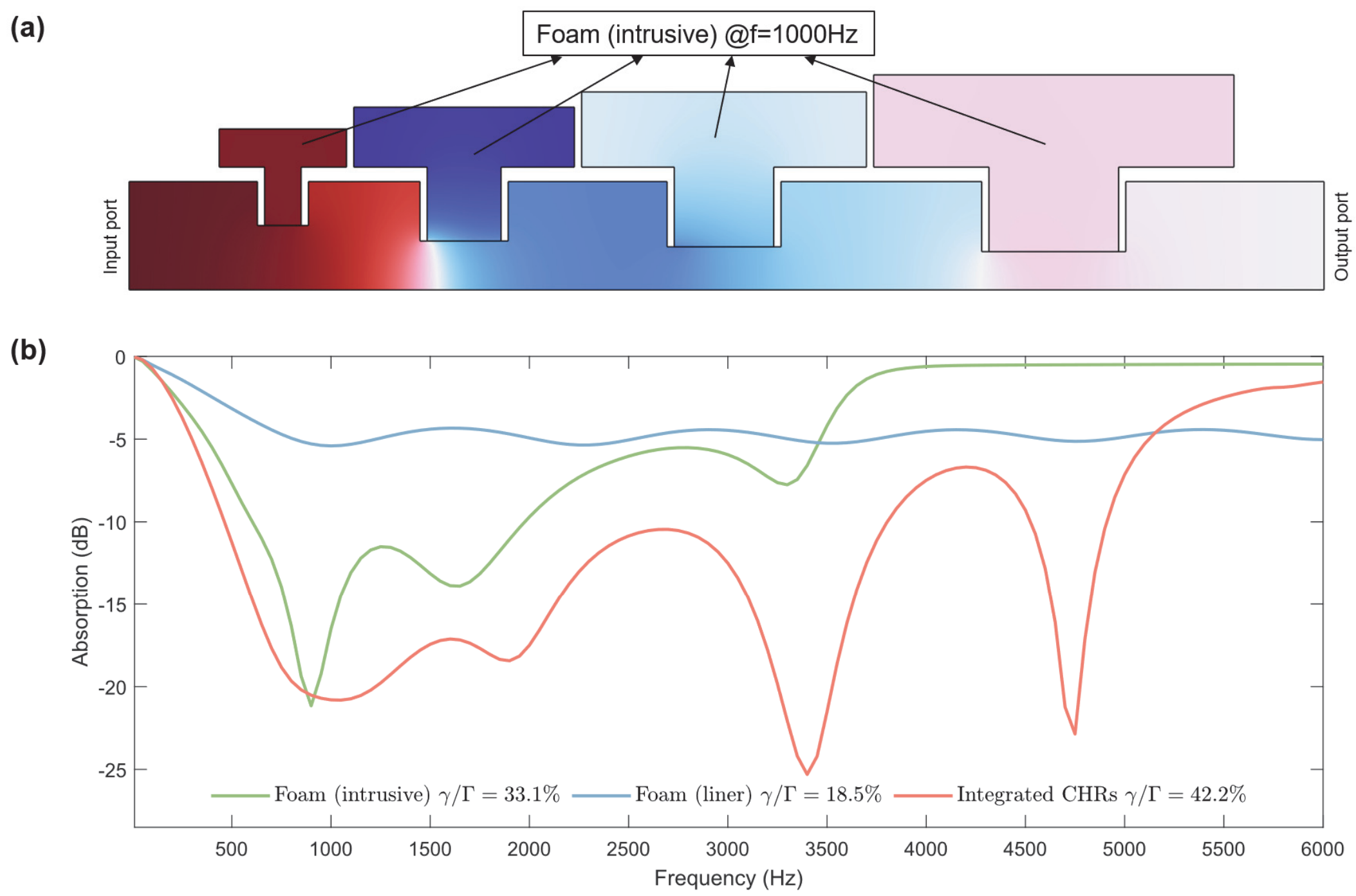

**Supplementary Figure 9. Additional simulation data for the intrusive foam. (a)** The geometric layout of the intrusive foam, which shares the same duct shape and volume as our integrated CHRs. The porous resonator domain was calculated using the same JCA parameters. **(b)** Absorption spectra are displayed on a dB scale: $10 \times \log_{10}(1 - A)$ for the intrusive foam, foam liner, and integrated CHRs. The legend shows the percentage between the full band integration of Eq. (12) ($\omega \to \infty$) and $\Gamma$.

## Supplementary References